\documentclass[12pt,preprint]{aastex}

\slugcomment{accepted by ApJ}

\shorttitle{LGS AO Search for Planetary Mass Companions}
\shortauthors{Biller et al.}

\usepackage{color}

\newcommand{\degs}{\mbox{$^{\circ}$}}
\newcommand{\Ks}{\mbox{$K_S$}}

\def\lesssim{\mathrel{\hbox{\rlap{\hbox{%
 \lower4pt\hbox{$\sim$}}}\hbox{$<$}}}}
\def\gtrsim{\mathrel{\hbox{\rlap{\hbox{%
 \lower4pt\hbox{$\sim$}}}\hbox{$>$}}}}

\begin{document}


\title{A Keck LGS AO Search for Brown Dwarf and Planetary Mass Companions to Upper Scorpius Brown Dwarfs}


\author{Beth Biller\altaffilmark{1,2}}
\affil{Institute for Astronomy, University of Hawaii, 2680 Woodlawn Drive, Honolulu, HI, 96822}

\author{Katelyn Allers}
\affil{Department of Physics and Astronomy, 153 Olin Science, 
Bucknell University, Lewisburg, PA 17837}

\author{Michael Liu}
\affil{Institute for Astronomy, University of Hawaii, 2680 Woodlawn Drive, Honolulu, HI, 96822}

\author{Laird M. Close}
\affil{Steward Observatory, University of Arizona, 933 N. Cherry
  Ave., Tucson, AZ 85721}

\and

\author{Trent Dupuy\altaffilmark{2,3}}
\affil{Institute for Astronomy, University of Hawaii, 2680 Woodlawn Drive, Honolulu, HI, 96822}

\altaffiltext{1}{now at: Max-Planck-Institut f\"ur Astronomie, K\"onigstuhl 17,
  D-69117 Heidelberg, Germany}
\altaffiltext{2}{Hubble Fellow}
\altaffiltext{3}{now at: Harvard-Smithsonian Center for Astrophysics, 60 Garden St., Cambridge, MA 02138}



\begin{abstract}

We searched for binary companions to 20 young 
brown dwarfs in the Upper
Scorpius association (145 pc, 5 Myr, nearest OB association)
with the the Laser Guide Star adaptive optics system
and the facility infrared camera NIRC2
on the 10 m Keck II telescope.  
We discovered a 0.14$\arcsec$ companion (20.9$\pm$0.4 AU) 
to the $<$0.1 M$_{\odot}$ object SCH J16091837-20073523. 
From spectral deconvolution of integrated-light near-IR spectroscopy
of SCH1609 using the SpeX spectrograph (Rayner et al. 2003),
we estimate primary and secondary spectral types of 
M6$\pm$0.5 and M7$\pm$1.0, corresponding to masses of
79$\pm$17 M$_{Jup}$ and 55$\pm$25 M$_{Jup}$ 
at an age of 5 Myr and masses of 84$\pm$15 M$_{Jup}$ and 
60$\pm$25 M$_{Jup}$ at an age of 10 Myr.
For our survey objects with spectral types 
later than M8, we find an upper limit on the binary
fraction of $<$9$\%$ (1-$\sigma$) at separations of 10 -- 500 AU. 
We combine the results of our survey with previous surveys of Upper
Sco and similar young regions to set the strongest constraints to date 
on binary fraction for young substellar objects and very low mass stars. 
The binary fraction for low mass ($<$40 M$_{Jup}$) brown dwarfs in Upper Sco
is similar to that for T dwarfs in the field; for higher mass brown
dwarfs and very low mass stars, there is an excess of 
medium-separation (10-50 AU
projected separation) young binaries with respect to the field.  These 
medium separation binaries will likely survive to late ages.
\end{abstract}





\section{Introduction}
Numerous brown dwarf binaries have been discovered in the the field
\citep{clo03,bur03,bou03,bur06,liu06}. 
Almost all of these have projected 
separations of $<$15 AU, with the majority having
separations of $<$7 AU.  This tight binary distribution was
initially viewed as evidence for the ejection scenario of brown dwarf
formation \citep{clo03}.  In the ejection scenario,
brown dwarfs are stellar embryos which are expelled from their natal 
subclusters due to interaction with other subcluster members,
therefore cutting off accretion.  Only tight brown
dwarf binaries can survive an ejection event \citep{rei01}.  

In the last decade a population of wide ($>$15 AU separation) 
very low mass star, brown dwarf, and
``planetary mass'' ($<$13 M$_{Jup}$) binaries have been discovered in young ($<$12 Myr) nearby 
clusters  \citep[][~see~Table~1~for~a~list~of~all~young~$\leq$0.1~M$_{\odot}$~binaries]{luh04, cha05, kra05, kra06, all06t, jay06,
clo07, kon07, tod10,bej08}.  These recent results suggest that the multiplicity properties of
young ($\sim$few Myr) substellar objects in star-forming regions
may be substantially different from the old ($\sim$few Gyr) field
population.  If common, these young binaries also provide serious
constraints for current theories of brown dwarf formation, since such
wide binaries cannot be formed by a non-dissipative ejection model \citep{bat09}.
However, most of these objects were either discovered serendipitously,
 are from surveys with unpublished statistics, or are from surveys
 with very few objects of comparable mass, 
so it is unknown how significant a population they 
form.   Here, we conduct a systematic 
survey to search for such binaries in Upper Sco, the nearest OB
association to the Earth.

\section{Sample Selection}

At an age of $\sim$5 Myr and a distance of 145 pc \citep{pre02}, the
Upper Scorpius OB association is one of the nearest sites of ongoing
star-formation and is intermediate in age between very young
star-forming regions
 such as Taurus ($<$1 Myr) and somewhat older young 
field objects ($\sim$100
Myr).  Additionally, Upper Sco is denser than nearby T 
associations such as Taurus and Chamaeleon but considerably
less dense than high-mass star-forming regions such as the 
Trapezium in Orion \citep{pre08}.
Binarity of young objects may vary both
as a function of age and environment \citep{pre08}. 
Certainly, the existence of very young, wide binaries 
in $<$2 Myr star-forming regions \citep[e.g.~][]{luh04, all06t, jay06,
clo07, kon07, tod10}
and the absence of such binaries in the field population
\citep[e.g.~][]{clo03} suggests some
evolution of brown dwarf binary properties 
must occur as a function of age.
Thus, Upper Sco provides a key
binarity data point, intermediate in both age and density.

Some low mass stars and high mass brown dwarfs in Upper Sco have
already been studied for binarity \citep{kra05, kra08}.  
Numerous binarity studies have been conducted which are sensitive
to very low substellar mass companions for very young clusters such as 
Taurus \citep{kra06,
  kon07}, Chamaeleon \citep{ahm07}, IC 348 \citep{luh05}, NGC 1333
\citep{gre07},
as well as Upper Sco \citep{kra05} -- however, the least massive
primaries observed in these surveys have generally
been limited to higher mass brown dwarfs ($>$40 M$_{Jup}$).
To date, only 18 objects with estimated masses $<$40 M$_{Jup}$ possess
AO or space based observations for binarity
which are published in surveys with well-defined contrast limits
(7 objects from Kraus et al. 2005, 11 objects from Luhman et al. 2005).

Here, we extend binarity results to lower
mass brown dwarfs and planetary mass objects in Upper Sco.
We surveyed a sample of
20 substellar objects in Upper Sco with reported spectral types of
M7.5 or later.  These objects were selected from those with
spectroscopic confirmation \citep{lod08} from the near-IR photometric and
proper motion surveys of \citet{lod07, sle08}.  According to the
models of \citet{bar03}, these objects are all substellar.  Indeed,
these are the least massive objects currently known in Upper Sco, 
with estimated masses of $<$40 M$_{Jup}$, thus this survey doubles the
number of young, low mass brown dwarfs imaged to search for binarity.  
At the 5 Myr age of Upper Sco, these objects are quite hot,
hence their late M and early L spectral types.  Eventually, these
objects will cool to become T dwarfs.  We focus in
particular on 18 objects selected from \citet{lod08} which form a
consistently selected and analyzed sample.  Survey objects are listed
in Table~\ref{tab:objects}.

\section{Observations and Data Reduction}


We observed 20 objects with the facility infrared camera NIRC2
and the Laser Guide Star adaptive optics system \citep{bou04, wiz04}
on the 10 m Keck II telescope.  Observations were conducted on the
nights of 2007-07-17, 2008-07-27, 2009-05-29, 2009-05-30, and
2009-06-30.  Conditions varied considerably between nights.  We used
the NIRC2 narrow camera, with a 9.963$\pm$0.005~mas~pixel$^{-1}$
platescale and
a 10.2$\arcsec$$\times$10.2$\arcsec$ field of view.  Search
observations were conducted in the Ks filter 
($\lambda_{central}$ = 2.146 $\mu$m).  Objects
were observed using a 3 point dither pattern, with a dither of
1-2.5$\arcsec$ between positions.  Observations are detailed in
Table~\ref{tab:observations}.  The data were reduced in real time at
the telescope using a custom IRAF pipeline.  In cases where
a candidate companion was detected at the telescope, immediate
followup observations in J and H bands were then conducted.  The observed
object FWHM ($K_S$ band) varied from 55 - 130 mas, with 
Strehl ratios in $K_S$ varying from 6 to 29$\%$.  FWHMs and Strehl
ratios were calculated using the standard Keck LGS routine nirc2strehl.pro.
Most objects  
appeared slightly elongated in the direction of the tip-tilt reference
star.  We used a custom IDL pipeline for a final reduction of the
data.  The IDL pipeline corrects for on-chip distortion,
flat fields, sky subtracts, and registers images using a
cross-correlation algorithm.
 



\section{Candidate Selection Technique and Tentative Companion Candidates}

Images were visually inspected for candidate companions.  A number of
faint candidate companions to several survey objects were identified at
separations of $>$1$\arcsec$.  To be considered true companions,
candidates must possess red colors similar to their primary and have
common proper motion.  Candidates
to USco J160603.75-221930.0 and USco J160723.82-221102.0 were
reobserved 1 year after the initial observations and found to be
background (i.e. not common proper motion objects).  Colors for other
candidates were checked in the ZYJHK bands of the UK Infrared Deep Sky
Survey (UKIDSS) or in the Digital Sky Survey (DSS).  The UKIDSS
project is defined in Lawrence et al (2007). UKIDSS uses the UKIRT
Wide Field Camera (WFCAM; Casali et al, 2007). The photometric system
is described in Hewett et al (2006), and the calibration is described
in Hodgkin et al. (2009). The pipeline processing and science archive
are described in Irwin et al (2009, in prep) and Hambly et al
(2008).  One candidate companion to USco J160830.49-233511.0 
was too faint to be detected in 
the UKIDSS data ($K_s\sim$19).  This candidate will be reobserved 
at Keck in Spring 2011; however, given its faintness and wide separation 
($\sim$5.4\arcsec, $\sim$780 AU), it is most likely background.  
All of the other $>$1\arcsec candidates were detected with S/N $>$ 10
in our Keck LGS AO data and were well-detected in the UKIDSS data.
All UKIDSS detected objects 
were found to have colors significantly bluer than their primary;
this is a clear sign that these objects are blue background objects as
opposed to a red brown dwarf or planetary mass companion.  


Since our survey objects are quite faint and our AO correction is in
general moderate, we did not attempt PSF subtraction to search for
faint companions within 1$\arcsec$ of the object.  Most of our objects
show some elongation towards the tip-tilt star.  Additionally, image
quality varied considerably between nights and during individual nights.
Thus, it was not possible to build a reliable synthetic PSF for PSF
subtraction.  We note that our brighter targets had a
number of superspeckles evident within 0.5$\arcsec$ of the primary
which can mimic the appearance of a companion. However, these
superspeckles modulate with wavelength and also evolve as a function
of time.  By comparing multiple images taken at different times or
wavelengths, it is almost always possible to distinguish speckles from
real companions.  In fact, we did initially flag a number of close-in
candidates which proved to be speckles upon further examination.

\section{Discovery of a Brown Dwarf  Companion to SCH 16091837-20073523}

A close candidate companion (0.14$\arcsec$) was detected around SCH
J16091837-20073523 (henceforth SCH 1609-2009) 
with colors consistent with a young substellar
object.  Photometry and astrometry for this object is presented in
Table~\ref{tab:objphot}.  Photometry and astrometry were determined
using two different methods: 1) DAOPHOT psf-fitting photometry using
IRAF and 2) synthetic psf-building photometry using BINFIT and
StarFinder in IDL.  

For the DAOPHOT psf-fitting photometry, 
a background object in the field 
(with separation $>$5$\arcsec$ from the primary and
blue colors as expected for a background object) was used as a PSF for
the allstar task.  The PSF object was well detected in both H and K bands.
However, our reduced AO correction in the blue J band relative
to the H and K resulted in a lower signal to noise detection of the
PSF star in the J band, diminishing our photometric accuracy in J
band.      

Since we were somewhat concerned that the background object used as 
a PSF at $>$5$\arcsec$ might be affected by anisoplanetism,
we also determined the 
the binary separation, position angle (PA), and flux ratio 
using the StarFinder PSF-fitting software package
\citep{dio00} as an independent confirmation of the 
DAOPHOT results. StarFinder simultaneously solves for an
empirical PSF model and the positions and fluxes of the binary
components.  Our $J$-band images had significantly poorer FWHMs and
Strehl ratios such that StarFinder did not converge on a solution.
Thus, for $J$ band we instead used a three-component Gaussian model,
as described by \citet{dup09}, to derive the binary
parameters from PSF-fitting.  The
uncertainties were determined from the rms of the best-fit parameters
for our individual dithered images.  We adopted the astrometric
calibration of \citet{ghe08}, with a pixel scale of
9.963$\pm$0.005~mas~pixel$^{-1}$ and an orientation for the detector's
$+y$-axis of $+$0$\fdg$13$\pm$0$\fdg$02 east of north.  We applied the
distortion correction developed by B. Cameron (2007, private
communication), which changed our astrometry below the 1$\sigma$
level.  Both DAOPHOT and StarFinder methods yielded nearly identical values (within the
cited errors) for photometry and astrometry.  

\begin{figure}
\epsscale{0.6}
\begin{tabular}[]{ccc}
\includegraphics[width=3in]{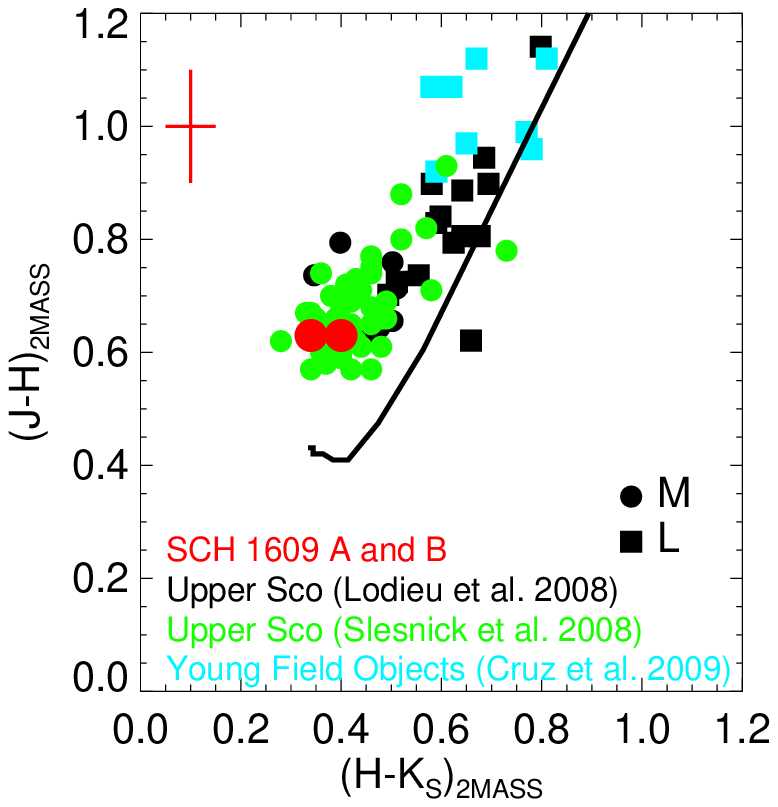}
& \includegraphics[width=3in]{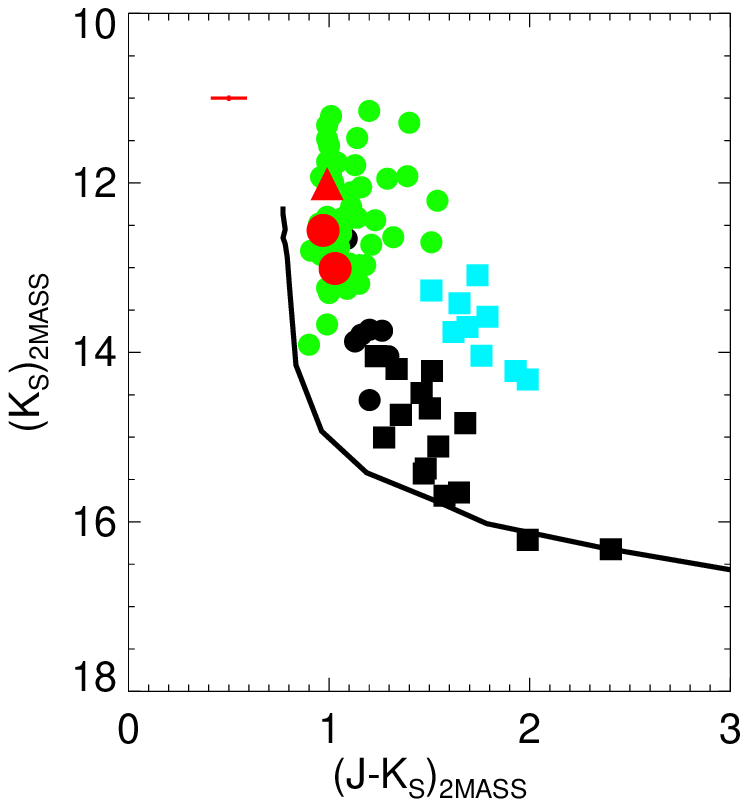}
 \\
\end{tabular}
\caption{Left: the $JH\Ks$ colors of SCH 1609 AB compared to Upper Sco 
objects with M and L spectral types (Slesnick et al. 2008, Lodieu et
al. 2008)
and young field brown dwarfs from Cruz et al. 2009.
SCH 1609ABs' colors are plotted as a red circles and are
  consistent with those of a mid to late M dwarf. 
Errors on SCH 1609AB photometry are shown in the top left corner.
The DUSTY 5 Myr isochrone is plotted as a solid line.
DUSTY models predict considerably bluer colors at these ages than is observed.
 Right: J-$K_{S}$ vs. K$_{S}$ for SCH 1609AB and the same set of 
comparison objects.
  SCH 1609AB are plotted as red circles; combined photometry for the 
system is plotted as a red triangle.
The DUSTY 5 Myr isochrone is again plotted as a solid line;
while $K_{S}$ band magnitudes agree with DUSTY predictions, colors are 
considerably redder than the predictions for M dwarfs.
  \label{fig:colors}
}
\end{figure}

We show the primary and companion on color-magnitude and 
color-color diagrams in Fig~\ref{fig:colors}.
The companion possesses very similar colors to its primary,
suggesting that it is a true substellar companion.  
We estimated the likelihood that this companion is a background object
using source counts from the 2MASS survey.  Within 1 degree of the
primary, 2MASS detects 527 objects with J of 13 mag or brighter, 506
objects with H of 12.4 or brighter, and 427 with K of 12 or brighter.
Thus, adopting the approach of \citet{bra00}, in particular, their
equation 1, we estimate the probability of finding an unrelated source
at least as bright as the observed companion within 0.14$\arcsec$ of
the primary to be $\sim$2.6$\times$10$^{-6}$.  

SCH 1609-2007 was reobserved with NIRC2 at Keck II on 1 May 2010.
The overall quality of the dataset was poor; however, we acquired 
sufficient data to demonstrate that the companion likely
has common proper motion with the primary.  
Measuring centroid positions of the primary and companion 
(as the 2nd epoch data were
 not high enough quality for psf-fitting photometry), 
the companion moved by $<$0.7 pixels relative to the primary between epochs, 
consistent with the errors in our simple center-of-light 
centroiding.    As no directly measured 
proper motion is available for SCH 1609-2007 we adopt the 
mean value of (-11, -25) mas yr$^{-1}$ 
for Upper Sco here \citep{deb97, pre98}.
At a distance of 145 pc, parallax motion for Upper Sco is quite small
-- $\sim$7 mas.  
As the parallax factor in the 2nd epoch observation was
similar to that in the first, we neglect parallax here. 
With the $\sim$10 mas pixel scale of
narrow camera, we would have expected the companion to move $\sim$2.3
pixels relative to the primary between epochs if it was a background
object at a much larger distance.  Thus, this is likely a proper motion pair.

\subsection{Spectroscopy and Spectral Type Estimates}

It has been noted that objects from \citet{sle08} are
considerably brighter than objects of similar spectral type from
\citet{lod08}.  In fact, in some cases, the discrepancy is as much as
2 or 3 magnitudes, e.g. the M8 objects SCH J1622-1951 
and USco J155419.99-213543.1
in the sample for this survey.  One possible explanation for this
discrepancy is that the later type \citet{sle08} objects consist
primarily of nearly equal mass binaries, such as SCH J1609
and likely SCH 1622-1951 as well.  However,
even after accounting for binarity, SCH J1609-2007 is still 2-3 mag
overluminous compared to similar objects from the \citet{lod08}
sample. 

\begin{figure}
\includegraphics[angle=90, scale=0.6]{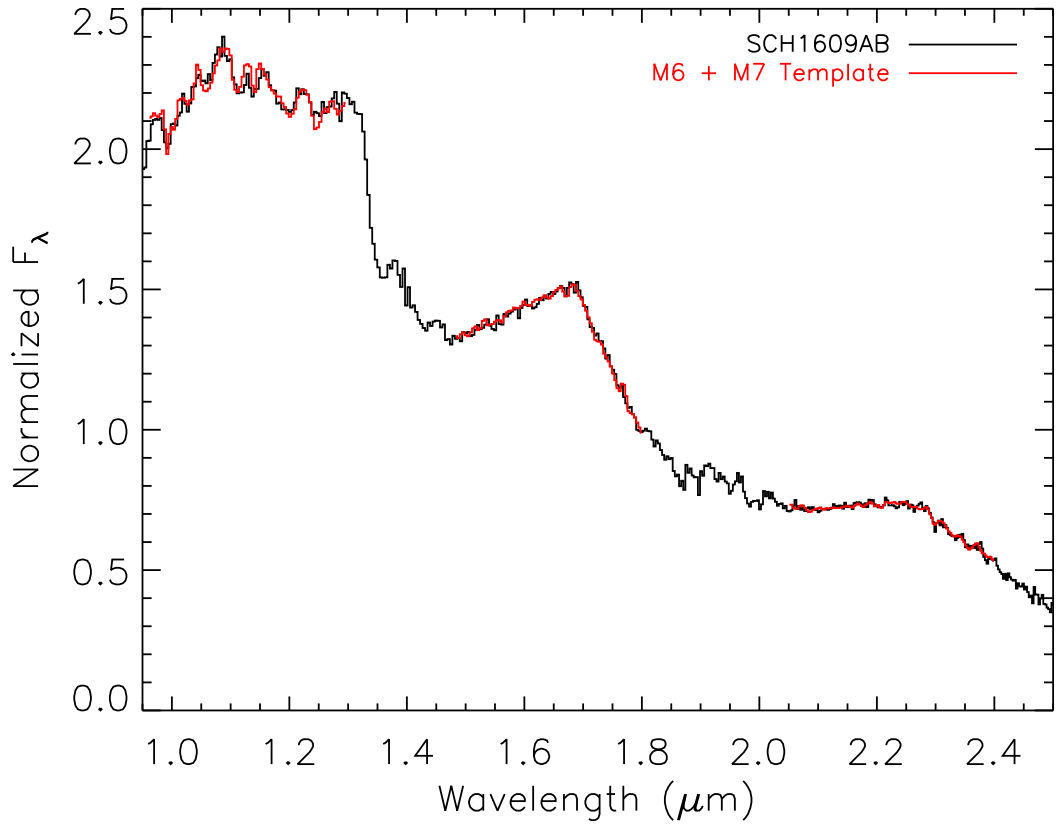}
\caption{Composite near-IR spectrum of SCH1609AB (black), compared to the
best-fitting synthetic composite spectrum (red). The synthetic
composite spectrum is the combination of UScoCTIO~75
(M6, Ardila et al. 2000; Preibisch et al. 2002) and
DENIS-P~J155605.0-210646 (M7; Martin et al. 2004, Slesnick et
al. 2008).
\label{fig:spectrum}}
\end{figure}

The discrepancy may also be due to systematic differences between
optical and infrared spectral types for these objects, which are right 
at the M to L type spectral transition.
All of the \citet{sle08} sources have optical spectral types
while the \citet{lod08} sources have infrared ones, so in 
effect we may be comparing apples vs. oranges.  Thus, to further
constrain the near-IR spectral type (and hence mass) of SCH J1609-2007AB   
we obtained integrated-light near-IR spectroscopy
of SCH1609-2007 on 2010 September 14 (UT) 
using the SpeX spectrograph \citep{ray03} on the NASA Infrared
Telescope Facility.  A series of 12 exposures of 30
seconds each were taken, nodding along the slit, for a total
integration time of six minutes.  Our observations were taken at an
airmass of 1.57, and the seeing recorded by the IRTF was 0$\farcs$9
The data were taken using the Low-Res prism with the 0$\farcs$5 slit
aligned with the parallactic angle, producing a 0.8--2.5~$\mu$m
spectrum with a resolution (R=$\lambda/\Delta\lambda$) of $\sim$150.
For telluric correction of our SCH1609-2007 spectrum, we observed a nearby
A0V star, HD~149827 and obtained calibration frames (flats and arcs).
The spectra were reduced using the facility reduction pipeline,
Spextool \citep{cus04}, which includes a correction for
telluric absorption following the method described in \citet{vac03}.  
Spectra and spectral fits are presented in 
Fig~\ref{fig:spectrum}.

SCH1609-2007AB was assigned a composite optical spectral type of M7.5 by
\citet{sle08}.  We determine spectral types for each
component by comparing our integrated light near-IR spectrum of
SCH1609-2007AB to synthetic composites generated from template near-IR
spectra of known members of Upper Scorpius (also taken with 
SpeX at the IRTF, at the same resolution as the SCH1609-2007AB
spectrum, Brendan Bowler, priv comm).  
Our Upper Scorpius templates have optical spectral types
ranging from M4 to M8.5.  We verified that our templates have near-IR
spectral types (calculated using the H$_2$O index of Allers et
al. 2007) that agree to within 1 subtype with their optical types.

To create our synthetic composite spectra, we first calculated
synthetic photometry for each template using the $J, H,$ and $Ks$ filter
profiles for NIRC2 and normalized each spectrum by the photometric
flux density.  We interpolated the templates to the same wavelength
grid as our spectrum of SCH1609-2007, and summed pairs of template spectra
together, multiplying the later spectral type template by the flux
ratio of the binary.  Following the technique described in Cushing et
al. (2008), we determined a multiplicative constant for each template
in each band ($J,H,$ and $Ks$), and computed a reduced $\chi^2$
performed over the wavelength ranges $\lambda$ = 0.96--1.3 $\mu$m,
1.48--1.8 $\mu$m, and 2.05--2.4$\mu$m.  The best fitting template is
the composite spectrum of UScoCTIO~75 (M6, Ardila et al. 2000;
Preibisch et al. 2002) and DENIS-P~J155605.0-210646 (M7; Mart\'in et
al. 2004, Slesnick et al. 2008).  We assign spectral types of M6.0
$\pm$ 0.5 to SCH1609-2007A and M7.0 $\pm$1.0 to SCH1609-2007B, where
uncertainties are determined from the spectral types of synthetic
composite spectra where $\chi^2 \ge \chi^2_{min}$ + 1, significantly 
earlier than the combined M7.5 spectral type from \citet{sle08}. 

\subsection{Mass Estimates}

We estimate the masses and effective temperatures 
of SCH 1609-2007 AB based on the DUSTY
models of \citet{cha00} and the temperature scale of \citet{luh04}.  The age 
of Upper Scorpius has been measured as 5 Myr, with a spread of up to 
2 \citep{pre99, sle08}, but more recent work suggests ages as old as 
10 Myr (Eric Mamajek, private communication). 

Thus to account for age spread, we estimate masses at discrete ages of
5 and 10 Myr using the DUSTY models \citep{cha00} and at an age range
of 5$\pm$1 Myr using dust-free models from the same group 
\citep{bar98, bar02}.\footnote[1]{While these two sets of models differ in colors due to different 
atmospheric compositions (dust grains or the lack thereof), they 
produce the same bolometric luminosities and effective temperatures
as a function of age and mass \citep{bar02}.  In particular, since
isochrones are only defined at 1, 5, and 10 Myr for the DUSTY models
as opposed to a much denser grid of isochrones for the 
\citet{bar98} models -- and the authors suggest caution using 
isochrones with ages $\leq$1 Myr,  
we have chosen to interpolate from the \citet{bar98} models when 
deriving masses at a range of ages to avoid inaccuracies from
interpolating from 1 Myr isochrones.}  For single age mass estimates,
we simulated the spectral type range of each object
with an input distribution of 10$^{6}$ Gaussian-distributed spectral type
values centered on the measured spectral type and with $\sigma$ set 
to the error on the measured spectral type.  We then converted
this input distribution to effective temperatures using the 
temperature scale of \citet{luh04} and to estimated mass using the 
\citet{cha00} models.  The estimated mass of each 
object was set to the mean
of the output distribution and the error on the mass was set to the 
standard deviation of the output distribution.  Via this method, 
we estimate primary and secondary 
masses of 79$\pm$17 M$_{Jup}$ and 55$\pm$25 M$_{Jup}$ 
at an age of 5 Myr and masses of 84$\pm$15 M$_{Jup}$ and 
60$\pm$25 M$_{Jup}$ at an age of 10 Myr.

For mass estimates for a range of ages, we simulated input spectral 
type and ages with an input distribution of 3$\times$10$^{4}$ 
Gaussian-distributed spectral type and age values.  As before, 
the spectral type values were centered on the  
measured spectral type, with $\sigma$ set 
to the error on the measured spectral type.  The center of the age
distribution was set as 5 Myr, with $\sigma$=1 Myr.
Interpolation with the \citet{luh04} temperature scale and 
\citet{bar98,bar02} models was performed to convert from spectral type
to estimated mass for the distribution.  Since the output 
distribution is somewhat asymmetric, we adopt the median as the 
best mass estimate and again use the standard deviation to set the 
error.  Via this method, we estimate primary and secondary masses
of 79$\pm$21 M$_{Jup}$ and 60$\pm$31 M$_{Jup}$.
Thus, for the range of ages that are realistic for this binary, the 
uncertainty in the measured spectral type dominates the mass estimate 
above and beyond any uncertainty in the age.

\subsection{Orbital Period Estimates}

We estimate the semimajor axis of SCH 1609-2007AB's orbit from its observed
separation. Assuming a uniform eccentricity distribution between 0 $<$ e
$<$ 1 and random viewing angles, Dupuy \& Liu (2010) compute a median
correction factor between projected separation and semimajor axis of
1.10$^{+0.91}_{-0.36}$  (68.3\% confidence limits).  
Using this, we derive a semimajor axis of 23.0$^{+19.0}_{-7.5}$ AU for
SCH 1609AB based on the observed separation in June 2009.
These correspond to an orbital period estimate of 310$^{+222}_{-211}$ yr,
for an assumed total system mass of 134$\pm$30 M$_{Jup}$.

\section{Achieved Contrasts and Limits on Minimum Detectable Companion Masses}

\begin{figure}
\includegraphics[angle=90, scale=0.5]{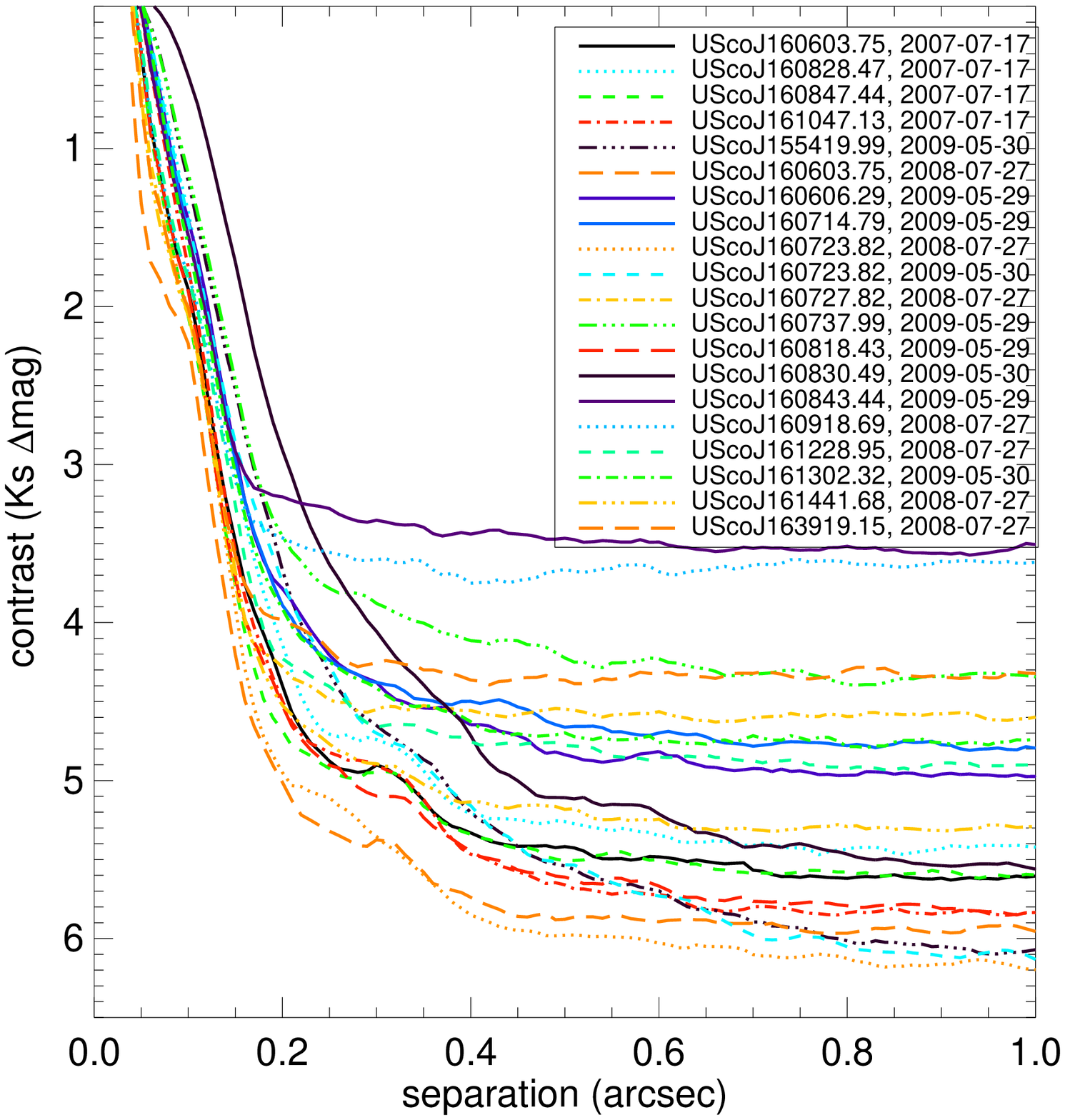}
\includegraphics[angle=90, scale=0.5]{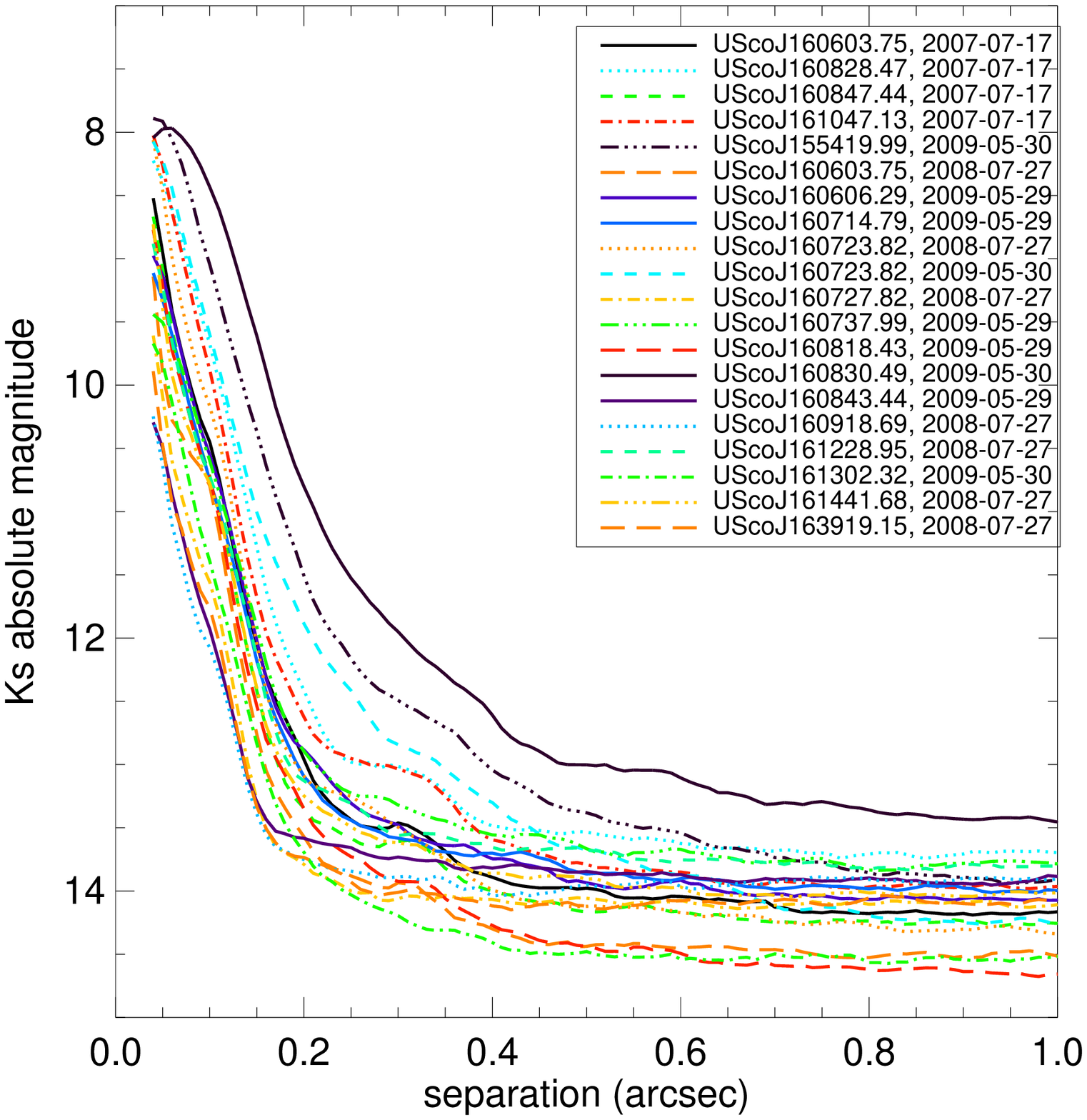}
\caption{Left: 5$\sigma$ contrast curves for 18 survey objects from \citet{lod08}.
Noise levels after data reduction were
calculated as a function of radius by calculating the standard
deviation in an annulus (with width equal to approximately the FWHM of
the PSF) centered on that radius.  Noise curves were then converted to
contrast in $\Delta$ magnitudes by dividing by the measured peak pixel
value of the object.  In general, we achieve contrasts of $>$4 mag at 
separations of $\geq$0.4'', sufficient to detect a 2MASS 1207 analogue 
at the distance of Upper Sco.
Right: Minimum detectable absolute magnitude for
the same 18 objects.  Contrasts were converted into absolute
magnitudes using photometry reported in \citet{lod08} and \citet{sle08},
and adopting a distance of 145 pc for Upper Sco.  A
filter transform was calculated from K to K$_s$ band using the spectra
from \citet{lod08} \label{fig:contrasts}}
\end{figure}

The 5-$\sigma$ contrast curves for 
our core sample of 18 objects from \citet{lod08} are presented in
Figure~\ref{fig:contrasts}.  Noise levels after data reduction were
calculated as a function of radius by calculating the standard
deviation in an annulus (with width equal to the FWHM of
the PSF) centered on that radius.  Noise curves were then converted to
contrast in $\Delta$mag by dividing by the measured peak pixel
value of the object.  Contrasts were converted into absolute
magnitudes using photometry reported in \citet{lod08}
and adopting a distance of 145 pc for Upper Sco.  A
filter transform was calculated from K to K$_s$ band using the spectra
from \citet{lod08}.  Absolute magnitudes of the faintest detectable
objects are also presented in Figure~\ref{fig:contrasts}.  A table of 
contrast values at separations of 0.07, 0.2, and 0.5\arcsec is 
presented in Table~\ref{tab:cont_q}.

To test the fidelity of our contrast curves, we inserted and retrieved
simulated objects in our data.  Objects were simulated as 2
dimensional gaussians with FWHMs from fits to the primary using the
IDL routine GAUSSFIT2D and contrasts from our measured contrast
curves.  Objects simulated with contrasts from our measured curves
were retrieved with S/N$\ge$5 for all survey targets down to
separations of 0.07$\arcsec$.  For separations down to 0.06$\arcsec$,
simulated objects were retrieved for half of our survey targets.  No
simulated objects were retrieved at separations $\le$ 0.05$\arcsec$.
Thus, we conclude that our measured contrast curves are a reliable
estimate of the detectable contrasts for potential companions down to
separations of 0.07$\arcsec$.

We note that these contrasts do not take into account confusion
between potential companions and speckles.  Our brighter targets had a
number of superspeckles evident within 0.5$\arcsec$ of the primary
which can mimic the appearance of a companion. However, these
superspeckles modulate with wavelength and also evolve as a function
of time.  By comparing multiple images taken at different times or
wavelengths, it is almost always possible to distinguish speckles from
real companions. 
Thus, since we can distinguish between the two, we believe that our
contrast curves adequately measure obtained contrasts for this survey,
despite potential speckle confusion.

\begin{figure}
\includegraphics[scale=0.7]{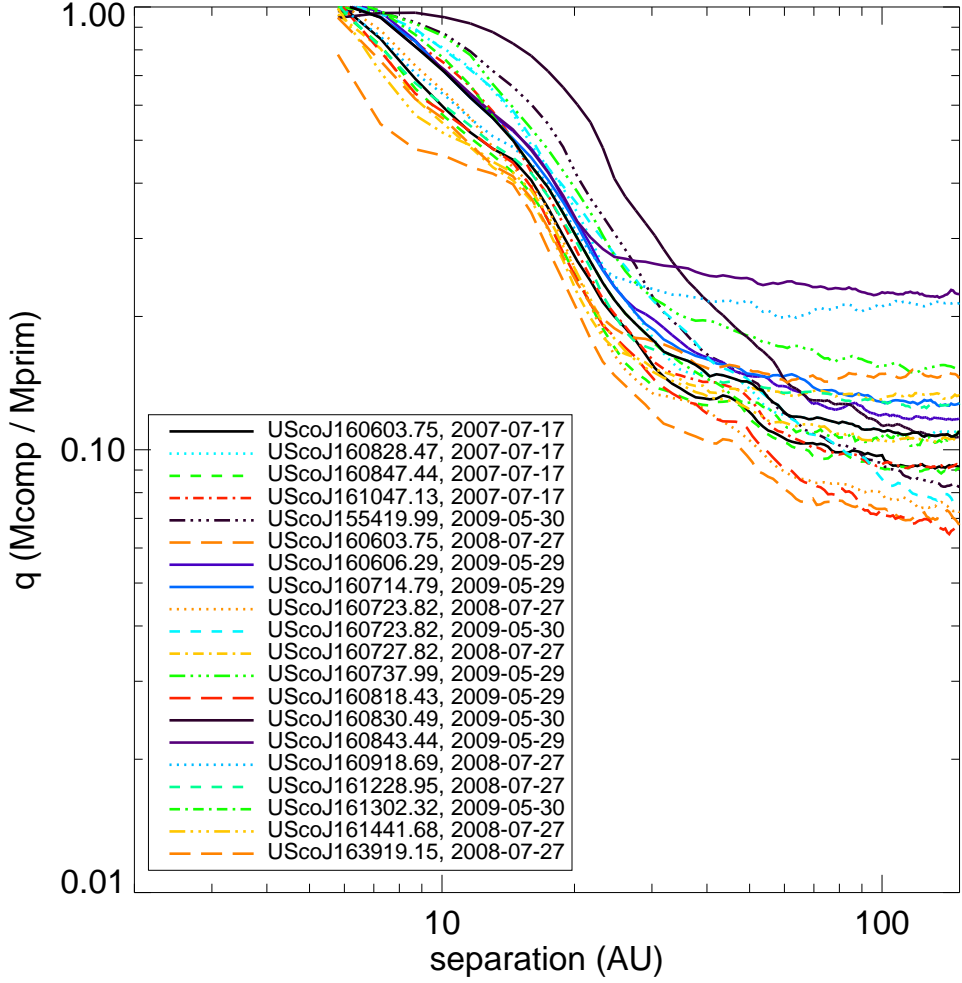}
\caption{Mass ratio (q) vs. separation using the DUSTY models. 
Contrasts were converted to minimum detectable mass ratios using the models
of \citet{cha00} at an adopted age of 5 Myr.   
For the best 75$\%$ of our data, we are 
complete to q$\sim$0.8 at 10 AU and complete to q$\sim$0.2 at
$\geq$20 AU.
\label{fig:q}}
\end{figure}

Contrasts were converted to minimum detectable mass ratios using the models
of \citet{cha00} at an adopted age of 5 Myr and assuming a similar 
bolometric correction (i.e. a similar spectral types for both objects)
between each target and any potential companion 
(Figure~\ref{fig:q} and Table~\ref{tab:cont_q}).  
We note that for the best 75$\%$ of our data
we are complete for all binaries with q$\geq$0.8 at 
separations $>$10 AU and all binaries with q$\geq$0.2 at separations
$>$50 AU.

\section{Discussion}

\subsection{Measured Binary Fraction}

We note that our newly discovered binary, SCH 1609AB, is
consistent with other young, wide very low mass binaries discovered,
with a wide ($>$10 AU) separation and nearly equal mass ratio (q$\sim$0.7).  
With only one companion detected as part of our survey, we cannot
place any new constraints on the mass ratio distribution or separation
distribution for young brown dwarf companions.
However, we have surveyed the largest sample to date of young brown
dwarfs with estimated masses $<$40 M$_{Jup}$ and can strengthen 
constraints on the binary fraction (10 -- 500 AU) of young objects in this
mass range.  We find an upper limit on the binary fraction (10 -- 500 AU)
of 9$\%$ (1-$\sigma$) for the 18 objects we surveyed from Lodieu et al. 2008
(calculated via the method of Burgasser et al. 2003).  
(We exclude the sources observed from the Slesnick et al. 2008
sample since they appear so much brighter than the Lodieu et
al. sources and likely have masses $>$ 40 M$_{Jup}$).

\subsection{Methods for Statistical Comparisons between Samples}

Given a sample of objects with true binary fraction $\epsilon_{bin}$,
the probability density of finding k binaries among n objects observed
is given by the binomial distribution:

\begin{equation}
f(k;n,\epsilon_{bin}) = \frac{n!}{k!(n-k)!}\epsilon_{bin}^{k}(1-\epsilon_{bin})^{n-k}
\end{equation}

In our case, we would like to invert this probability density in order 
to obtain the probability of the sample having 
a given binary fraction $\epsilon_{bin}$ in the case that we measure
k binaries among n objects.  To estimate the true binary fraction for
our sample, we can then derive a confidence interval
(presented as 1$-\sigma$ intervals here) around the maximum of this 
probability density in which we expect the true binary fraction to reside.
The probability density function and the resulting confidence
intervals can either be
calculated numerically (via e.g. the method of Burgasser et al. 2003) 
or by Bayesian posterior inference (see e.g. Sivia et al. 2006
and Cameron 2010).

Quantitatively 
comparing the binary fractions (with error bars included 
from confidence intervals) from sample to sample requires some
additional mathematical machinery. 
In some cases the confidence intervals overlap for binary fractions
derived for different samples  -- however, it is not immediately clear 
how statistically significant this correlation is. 
In comparing two samples of objects the
question we wish to answer is: are they drawn from the same binomial
distribution  with $\epsilon_{bin}$ or 
from different distributions?  To determine the likelihood
that two samples are drawn from the same binomial distribution, 
we adopted both 
the Fischer exact test method (used by \citet{ahm07} and described
in the appendix of \citet{bra06}) as well as a Bayesian approach,
derived below (derivation adopted from Carpenter 2009).


According to Bayes' Theorem:
\begin{equation}
prob(hypothesis;data,I) \propto prob(data;hypothesis, I) \times prob(hypothesis;I)
\end{equation}

where (in this case) ``I'' is prior information, ``data'' 
is our measured sample, and ``hypothesis'' is the hypothesis (e.g. in
this case we hypothesize that for brown dwarfs,  
phenomenon of binarity can be modeled as a binomial distribution with
binary probability $\epsilon_{bin}$).  
In Bayesian terms, prob(hypothesis;I) is the prior probability and
represents what we initially know regarding the truth or falseness of
the hypothesis while prob(data;hypothesis,I) is the likelihood
function and gives the likelihood of each possible experimental
outcome given the adopted model for the data.  Combining the two
gives prob(hypothesis;data,I), the posterior probability -- the
likelihood of a given model, in light of the measured data.

In this case, we would like to derive the posterior 
probability density not for 
each individual sample but for the difference of the two:

\begin{equation}
\delta = \epsilon_{bin1} - \epsilon_{bin2}
\end{equation}

To do this, we first must derive the posterior probability
distributions appropriate for each of the two binomial distributions
we are comparing.  Our likelihood function is again given by the binomial 
distribution, where $\epsilon_{bin}$ is the true binary fraction, 
for the sample, n is the total number of objects observed, and 
k is the number of binaries found:

\begin{equation}
prob(data;hypothesis,I) = f(k;n,p) = \frac{n!}{k!(n-k)!}p^{k}(1-p)^{n-k}
\end{equation}

For the prior probability, prob(hypothesis;I), we simply adopt a
uniform distribution from 0 to 1, i.e. the binary fraction 
must be between 0 and 1.  This can also be written in terms of the 
beta distribution, a special case of the Dirichlet distribution with 
only two parameters defined on the interval (0,1):

\begin{equation}
f(x;\alpha,\beta) =
\frac{\Gamma(\alpha+\beta)}{\Gamma(\alpha)\Gamma(\beta)}x^{\alpha-1}(1-x)^{\beta-1}
\end{equation}

Adopting $\alpha=\beta=$1, f(x;1,1) reduces to a uniform distribution,
thus:

\begin{equation}
prob(hypothesis;I) = 1~in~the~interval~0,1 = Beta(1,1)
\end{equation}

The advantage of choosing the Beta distribution to represent the 
prior probability is that the Beta distribution is a conjugate
distribution to the binomial distribution.  In other words, if the
prior probability is a Beta distribution and the likelihood is a
binomial distribution, then the posterior probability will also be 
a Beta distribution.  In this case, it is instructive to view the 
likelihood as an ``operator'' on the prior probability which 
produces as a result the posterior probability.  
When a binomial distribution ``operates'' on a beta distribution with
prior hyperparameters $\alpha$ and $\beta$, the result is the
following posterior distribution:

\begin{equation}
prob(hypothesis;data,I) = Beta(k+\alpha, n-k+\beta)
\end{equation}

Thus, in our case where $\alpha=\beta=$1:

\begin{equation}
prob(hypothesis;data,I) = Beta(k+1, n-k+1)
\end{equation}
 
Thus, our posterior probability distributions for each sample are
given by:

\begin{equation}
prob(\epsilon_{bin1};k_1, n_1) = Beta(\epsilon_{bin1};k_1+1, n_1 - k_1 + 1)
\end{equation}

\begin{equation}
prob(\epsilon_{bin2}|k_2, n_2) = Beta(\epsilon_{bin2}|k_2+1, n_2 - k_2 + 1)
\end{equation}

The posterior probability density for $\delta$ is then given by:

\begin{equation}
prob(\delta;k, n) = \int_{-\infty}^{\infty} \! Beta(\epsilon_{bin}|k_1+1, n_1 - k_1
+ 1) Beta(\epsilon_{bin}-\delta|k_2+1, n_2 - k_2 + 1) \, \mathrm{d}\epsilon_{bin}
\end{equation}

\begin{figure}
\includegraphics[angle=270, scale=0.3]{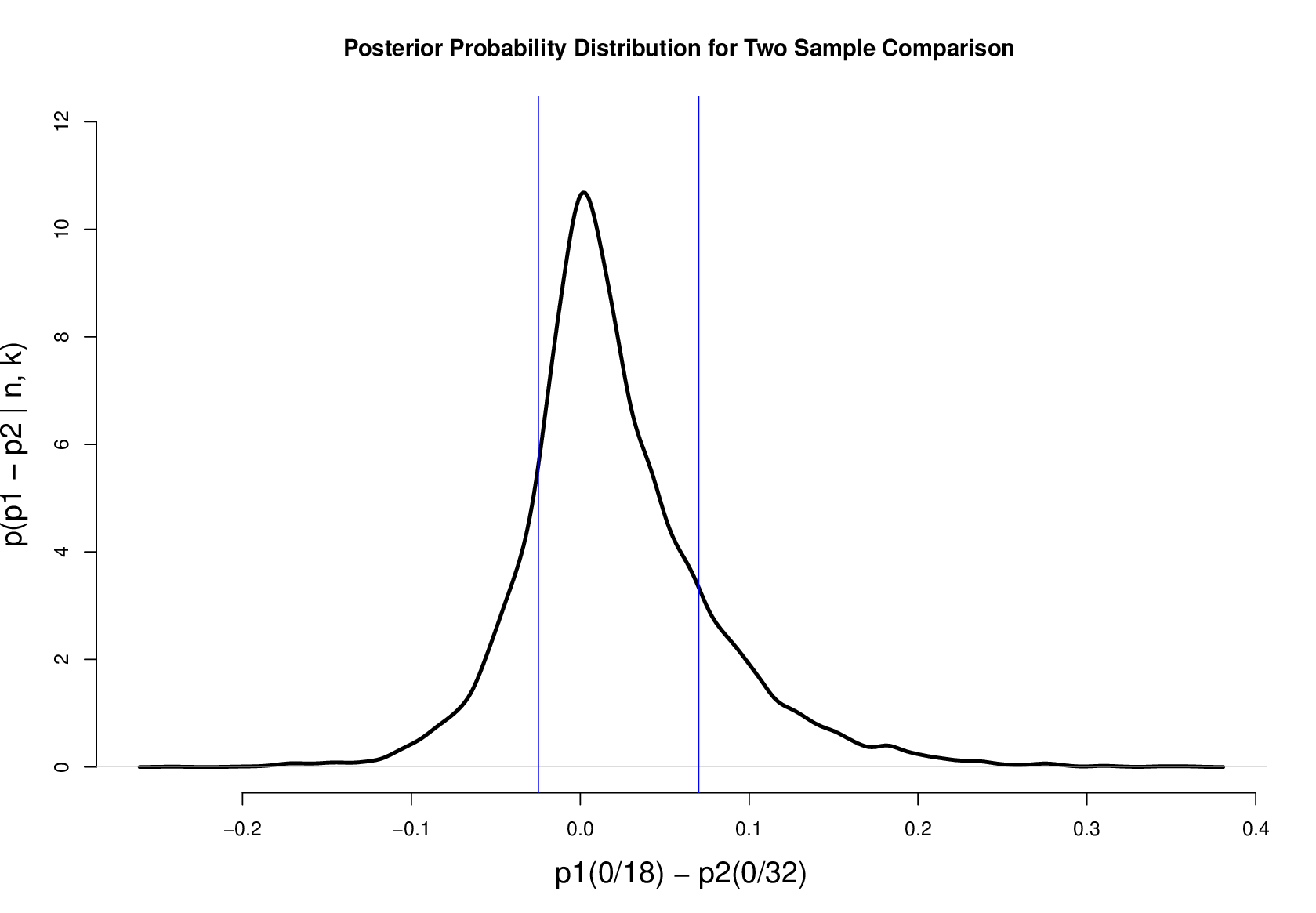}
\includegraphics[angle=270, scale=0.3]{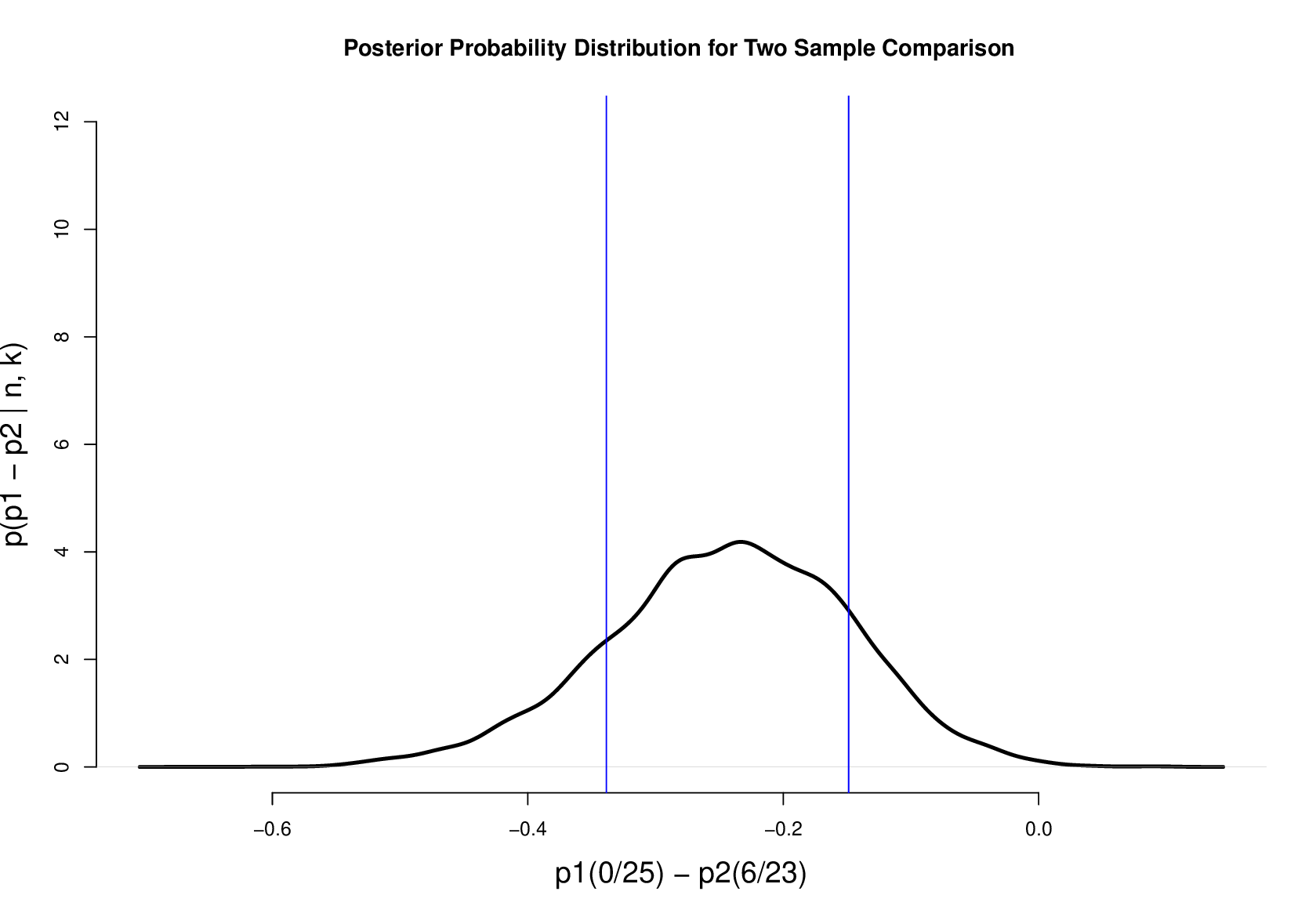}
\caption{Sample posterior probability distributions.  
Blue lines show the 1-$\sigma$ confidence intervals on 
$\epsilon_{bin1} - \epsilon_{bin2}$ = p1 - p2.  Left:
two samples which are likely 
drawn from the same binomial distribution 
(specifically $<$0.04 M$_{\odot}$ BDs in Upper Sco 
with 0 binaries detected out of 18 objects, compared with field 
T dwarfs, with 0 binaries detected out of 
32 objects).  The posterior probability 
distribution is strongly peaked at 0 and shows little spread.
Right: two samples which are likely 
drawn from different binomial distributions (specifically $<$0.04 M$_{\odot}$
Upper Sco, Taurus, and Chamaeleon objects, with
 0 binaries detected out of 
25 objects, compared with 0.07-0.1 M$_{\odot}$ Upper
Sco, Taurus, and Chamaeleon objects, 
with 6 binaries detected out of
23 objects).  The posterior probability distribution
in this case peaks considerably away from 0, and is wider and flatter 
than the previous case.  In particular, at the 1-$\sigma$ level  
$\epsilon_{bin2}$ = p2 for the 2nd distribution is between 
$\epsilon_{bin1}$ + 0.15 and $\epsilon_{bin1}$ + 0.33.
\label{fig:bayesian}}
\end{figure}

We used Monte Carlo methods in the R programming language 
to evaluate this integral. 10$^4$ random deviates were taken from each
posterior probability Beta distribution and the posterior probability 
density for $\delta$ was determined from these.  Two representative
posterior probability densities (for the case where both samples
likely share the same binomial distribution and also the case where
binomial distributions differ between samples) are presented in 
Fig.~\ref{fig:bayesian}.  Here we present the 
1-$\sigma$ (68$\%$) and 2-$\sigma$
(95$\%$) confidence intervals from the posterior probability density 
for $\delta$ as a counterpart to the
Fisher exact test likelihoods.  These confidence intervals quantify 
the probable relationship between the true binary fractions of the two 
samples.  For example, for the second comparison presented in 
Fig.~\ref{fig:bayesian}, a sample with 0 binaries detected out of 
25 objects compared with a sample of 6 binaries detected out of
23 objects, at the 1-$\sigma$ level  
$\epsilon_{bin2}$ for the 2nd distribution lays between 
$\epsilon_{bin1}$ + 0.15 and $\epsilon_{bin1}$ + 0.33.

\subsection{Brown Dwarf Binary Fraction as a Function of Mass}

We compare our measured binary fraction to that of more massive
brown dwarfs and very low mass stars in the Upper Sco embedded cluster.  
\citet{kra05} surveyed 12 brown dwarfs and very low mass stars
 with ACS on HST.  These
objects have estimated masses of 0.04 -- 0.1 M$_{\odot}$ and thus
comprise a higher mass sample than our survey.  \citet{kra05}
discovered three binaries in this sample, one of which (USco-109 AB) 
is below the sensitivity of our survey to detect, with a projected 
separation of only $\sim$5 AU.  Thus, for the
purposes of comparison, we adopt a binary fraction of 2/12 =   
17$_{-6}^{+15}$ \% 
for the \citet{kra05} sample.
The likelihood that the \citet{kra05} sample is drawn from the same 
distribution as ours is 0.15, with a 1$-\sigma$ Bayesian confidence 
interval of $\epsilon_{bin1} - \epsilon_{bin2}$  = -0.28, 0.15.
Thus, as noted by previous authors \citet{kra05}, 
the binary fraction in Upper Sco 
continues to decrease with decreasing primary mass.

This comparison is limited by the relatively 
small number of objects observed in Upper Sco.
Thus, to improve statistics, we have compiled a larger list using 
objects from similar surveys of other young, nearby regions -- specifically
Taurus \citep[$<$1~Myr,~145~pc,~objects~from][]{kra06, kon07} 
and Chamaeleon \citep[$<$3~Myr,~160~pc,~objects~from][]{ahm07}. 
We include only companions that would have been detected at the 
sensitivity level of our survey and initially limit this
analysis to nearby clusters ($<$200 pc) since more distant
clusters (e.g. NGC 1333, IC 348, Serpens) are more than 250 pc distant
and do not reach comparable sensitivity levels at 10 AU.
All selected surveys have similar sensitivity levels 
(complete to q$\sim$0.8 at 10 AU, complete to q$\sim$0.2 - 0.3 at
$\geq$20 AU) so it is unlikely that our survey would have discovered a
binary at a separation $>$10 AU missed by these other surveys, and vice versa. 
We adopt 3 mass bins
for this analysis: (1) high mass (0.07 -- 0.1 M$_{\odot}$), with
6 binaries detected out of 23 objects surveyed 
(6 objects from Ahmic et al. 2007, 
5 from Kraus et al. 2006, 6 from Kraus et al. 2005, 
4 from Konopacky et al. 2007, 
and the two objects from the Slesnick et al. 2008 sample from
the current survey), (2) medium mass (0.04 -- 0.07
M$_{\odot}$), with 0 binary detected out of 18 objects surveyed 
(4 objects from Ahmic et al. 2007, 6 from Kraus et al. 2005, 
and 8 from Kraus et al. 2006), and (3) low mass ($<$0.04 M$_{\odot}$), 
with 0 binaries detected out of 25 objects surveyed (7 objects 
from Kraus et al. 2006 and the 18
objects from the Lodieu et al. 2008 sample surveyed herein).
We note that while a number
of additional binaries are known in this mass range, e.g. 2MASS 1207AB 
\citep{cha05}, 2M 1622 \citep{all06t, all06, jay06, all07, clo07}, 
UScoCTIO 108 \citep{bej08}, and 2MASS 0441 \citep{tod10}, survey statistics
are not available for these objects and thus we cannot include them
in our sample.
Binary fractions and likelihoods
between bins as a function of mass are presented in Table~\ref{tab:tests1}.
As expected, the binary fraction decreases monotonically with primary
mass.  The likelihood that the low mass bin ($<$0.04 M$_{\odot}$)
objects share the same binary fraction as the high mass bin 
($>$0.07 M$_{\odot}$) is less than 0.02, with a 
1$-\sigma$ Bayesian confidence 
interval of $\epsilon_{bin1} - \epsilon_{bin2}$  = -0.34, -0.15.
 
\subsection{Brown Dwarf Binary Fraction as a Function of Age}

By ages of 1 Gyr, most of our survey objects will have cooled to become
T dwarfs.  Thus, it is interesting to compare the primordial binary
fraction of these objects to the binary fraction of similar objects in
the field.  Our survey is only sensitive to companions at projected
separations of $>$10 AU, however, this is a particularly interesting
separation space to probe, as older field T dwarf binaries rarely
have separations this large \citep{bur03,bur06}.  In fact, 
of the 32 T dwarfs surveyed in \citet{bur03, bur06}, no
companions were detected with separation $>$10 AU (down to
q$\geq$0.4, i.e. comparable sensitivity limits to our survey).
This places an upper limit on the binary fraction $>$10 AU of 5$\%$.
Again using the Fischer exact test method, we found a likelihood of
1 with a very tight 1$-\sigma$ Bayesian confidence 
interval of $\epsilon_{bin1} - \epsilon_{bin2}$  = -0.02, 0.07
-- i.e., given the small sizes of both of these samples, they are
very likely drawn from the same parent sample.  Thus, the very low mass 
brown dwarf binary fraction appears to be similar for both young and
field objects.  Binary fractions, likelihoods, and Bayesian confidence
intervals 
between bins as a function of age are presented in Table~\ref{tab:tests2}.

Do higher mass objects ($>$0.07 M$_{Sun}$) in young clusters also 
have a similar binary fraction ($>$10 AU) as their counterparts in the field?
We compare the binary fraction
($>$10 AU separation) for 6 binaries discovered out of 
23 young objects (the ``high mass'' bin from the previous section) 
drawn from binarity surveys of 
Upper Sco \citep{kra05}, Taurus \citep{kra06,kon07}, 
Chamaeleon \citep{ahm07} and this work with that of 1 binary ($>$10 AU 
separation) discovered from 39 field M8--L0.5 objects from
Close et al. 2003.  These two samples share a similar mass range (primary
mass between 0.07 -- 0.1 M$_{\odot}$), but very different wide binary fractions: 
26$^{+11}_{-7}$\% for the young sample vs. 2.6$^{+5.4}_{-0.06}$\%
for the old field sample.  
Using the Fischer exact test, the likelihood that these two samples
are drawn from the same binomial distribution is 0.01 
with 1$-\sigma$ Bayesian confidence 
interval of $\epsilon_{bin1} - \epsilon_{bin2}$  = 0.14, 0.32.
Thus, for objects with mass $>$ 0.07 M$_{\odot}$, 
there is an overabundance
of 10--50 AU separation very low mass binaries in young clusters 
relative to the field.

Upper Sco is a somewhat older and higher density OB association, while
Taurus and Chamaeleon are more diffuse, younger T clusters.
Thus, we also compare
binary fraction between these two different ages and environments.
Combining the sample described in the previous section and separating by region, we find
3 binaries detected from 34 objects in Taurus and Chamaeleon and 
3 binaries detected from 32 objects in Upper Sco. The binary fraction is
nearly the same between the two, although it is important to note that
the sample in Taurus and Chamaeleon has systematically higher primary
masses than
that in Upper Sco (dominated by the 18 very low mass brown dwarfs
surveyed in this paper.)  Thus, given the trend in binary fraction 
with mass, the binary fraction in Upper Sco may be considerably higher
than in Taurus.


\subsection{Trends in Very Wide Binarity (30 -- 500 AU) as a Function
  of Age and Mass}

We initially limited our statistical analysis
to nearby clusters ($<$200 pc) since more distant
clusters (e.g. NGC 1333, IC 348, Serpens) are more than 250 pc distant
and do not reach comparable sensitivity levels at separations of 10-30 AU.
However, including results from surveys of these more distant clusters
significantly boosts sample size.  In particular, including
the results from the \citet{luh05} HST survey of IC 348
(2 Myr, 315 pc) introduces 31 additional $\leq$0.1 M$_{\odot}$ objects
into this analysis.  In order to match the achieved 
contrast and physical resolution 
of the \citet{luh05} survey with those of nearer regions (Upper Sco, 
Taurus, Chamaeleon) we only consider results for separations from
30-500 AU. 

Including objects from \citet{luh05} in our 3 mass bins
from earlier sections, we now find: (1) in the high mass bin 
(0.07 -- 0.1 M$_{\odot}$), 
3 binaries are detected out of 43 objects surveyed 
(USco-55 and USco-66 from Kraus et al. 2005 and 
USco1609 from this work have separations $<$30 AU
and thus would not be detected at the combined sensitivity limits for 
our composite survey), (2) in the medium mass bin (0.04 -- 0.07
M$_{\odot}$), 0 binaries detected out of 20 objects surveyed, 
and (3) in the low mass bin ($<$0.04 M$_{\odot}$), 
0 binaries detected out of 36 objects surveyed. 

As before, the lowest mass cluster bin possesses a very similar upper
limit on binarity as the field T dwarf bin and wide binaries seem to
be rare in both the medium and low mass cluster bins.
Comparing the high mass bin with the low mass bin, we find a
likelihood of 0.25 that these two samples are drawn from the same 
binomial distribution, with a 1$-\sigma$ Bayesian confidence 
interval of $\epsilon_{bin1} - \epsilon_{bin2}$  = -0.11, -0.02.  
We also compare the high mass bin with the Close et al. 2003 sample 
(adjusting contrast levels, we now find 0 binaries imaged with
separations $>$30 AU out of 39 surveyed objects).  According to the
Fisher Exact Test, the likelihood 
that these two samples were drawn from the same binomial distribution
is 0.24 (as opposed to 0.01 for the same bin in the nearby sample.) 
Thus, a significant overdensity of young binaries relative to the
field is apparent in this sample only at moderate separations (10-30
AU), and not at wide (30-500 AU) separations.

\subsection{Stability of 10-50 AU Separation Binaries in Young Nearby
  Starforming Regions}

Up to $\sim$25$\%$ of very low mass (henceforth VLM) stars and 
substellar objects
 in young star forming regions may have 
companions at separations $>$10 AU.
However, very low mass star / 
brown dwarf binaries with separations $>$15 AU are rare in the field.
Of $\sim$100 VLM binaries compiled at vlmbinaries.org, only
$\sim$10$\%$ have separations $>$15 AU.
Assuming a binary fraction 
of $\sim$10$\%$, this means that 
less than 1$\%$ of field VLMs have companions at 
separations $>$15 AU \citet{clo07}.

\citet{clo07} suggest that very wide ($>$50 AU) young brown
dwarf binaries are disrupted within the first 10 Myr of their
existence by interactions with stars in their natal cluster.  To set 
limits on the survival time of a young wide binary in its natal
cluster, they adopt the 
analytic solution of Fokker-Plank (FP) coefficients from \citet{wei87}
which describes the advective diffusion of a binary due to stellar 
encounters.  Namely, from this solution, the time t$_{*}$ necessary to
evaporate a binary with initial semimajor axis a$_0$ is:

\begin{equation}
t_{\ast} \sim 3.6 \times 10^{5} (\frac{n_{\ast}}{0.05 pc^{-3}})(\frac{M_{tot}}{M_{\odot}}) (\frac{M_{\ast}}{M_{\odot}})^{-2}(\frac{V_{rel}}{20 km s^{-1}}) (\frac{a_{\circ}}{AU})^{-1} 
\end{equation} 
  
where n$_*$ is the number density of stellar perturbers of mass
M$_*$ and relative velocity V$_{rel}$.  Using this relationship,
\citet{clo07} determine that young wide VLM binaries such as 2M 1207-39AB
will not survive 10 Myr of interactions with 0.7 M$_{\odot}$ stellar
perturbers with a number density n$_*$ of 1000 pc$^{-3}$.  Thus, 
\citet{clo07} show that most of these binaries 
will not survive to join the
field if born in a dense starforming region. We determine
here whether the same is true for moderately wide 10 -- 50 AU binaries.
While \citet{clo07} assume a number density of 
nearby stars of 1000 pc$^{-3}$, which is appropriate near dense
core regions, it is probably too high for objects in diffuse areas of 
Taurus or Chamaeleon.  Assuming a typical density of 100 pc$^{-3}$ for
for Taurus, Ophiuchus, and Upper Sco, we repeat this calculation for the six
binaries that fall into our highest mass bin 
(specifically, CFHT-Tau 7, CFHT-Tau 17, and
CFHT-Tau 18 from \citet{kon07}, USco-55 and USco-66 from 
\citet{kra05}, and SCH 1609AB, the newly discovered binary presented herein).
We find that all of these binaries are quite stable 
and will survive $>$10 Myr in either
a 100 pc$^{-3}$ environment or a 1000 pc$^{-3}$ environment (i.e. 
long enough to join the field population).  An environment with 
stellar densities $>$10$^4$ pc$^{-3}$ (equivalent to the Trapezium
cluster in Orion) is necessary to disrupt these binaries on $<$10 
Myr timescales.   
 
The existence of a significant population of these medium-separation 
binaries 
 presents a conundrum, since
very low mass stars and brown dwarf binaries
with separations $>$15 AU are rare in the field.
However, the field brown dwarf population encompasses a mix of
objects which formed in a variety of different star-forming regions.
\citet{clo07} suggest that brown dwarf binaries with separations $>$20
AU are found rarely in the field because they can only form 
in low-density star-forming regions, while the majority of field
objects formed in denser initial regions where any such binary would be 
disrupted.  However, other authors have
suggested that most stars in the field likely form in OB associations
like Upper Sco \citep{kon07,pre08}, so this is problematic.

The existence of this population of moderately wide young 
brown dwarf binaries in lower density young clusters initially suggests 
that most (predominantly single) 
field brown dwarfs must form in high stellar density regions which
disrupt such wide binaries by late ages.
However, this supposition relies on our ability to distinguish
between 
``typical'' vs. ``atypical'' star-forming regions, as well as to truly
disentagle the primordial vs. evolved populations.  In other words, 
the evolved population is the outcome of the formation mechanism 
plus any subsequent evolution in the cluster.  Different combinations
of formation and subsequent evolution may form the same evolved population.
Here, we have placed constraints on binary evolution within a relatively
diffuse cluster environment; placing constraints on formation
mechanism is more difficult.

Forming brown dwarfs at all has always been a tricky prospect theoretically.
Brown dwarf formation theories require either: (1) a mechanism to
produce very low Jeans masses
\citep[e.g.~turbulent~fragmentation,~gravitational~fragmentation~of~infalling~gas,and~gravitational~
fragmentation~with~a~magnetic~field,~][]{pad04,bat09,bon08,pri08} 
or (2) a method to circumvent the need for very low Jeans masses,
\citep[e.g.~ejection,~or~gravitational~instability~followed~by~binary~disruption~][]{rei01,sta07,sta09} 
Unfortunately, more information is needed regarding the physical
properties of these star-forming regions (i.e. measurement of
turbulent motions, magnetic fields) to distinguish between these models.
For instance, widespread filamentary structure has recently been observed by
Herschel in very young star-forming clouds in Aquila and Polaris \citep{men10}.  However, it is not
currently clear what causes these filaments; if turbulence or magnetic
fields are the dominant cause, this has significant ramifications for
subsequent brown dwarf formation in these regions.

Likely a mix of formation mechanisms are at play in any given region,
the detailed physics of which may vary from region to region.
Disentangling these physics is a difficult
prospect and requires more information than just binary
fraction.  While we can rule out pure ejection (without any dissipation
from e.g.~a circumstellar disk) from our measured binary 
fraction and the existence of a significant population of $>$10
AU separation binaries, other models may produce a significant wide 
binary population which may be disrupted by late ages in dense clusters.


\section{Conclusions}

We searched for binary companions to 20 brown dwarfs in Upper
Scorpius (145 pc, 5 Myr, nearest OB association)
with the laser guide star adaptive optics system and 
the facility infrared camera NIRC2  
on the 10 m Keck II telescope.  This survey is the most extensive
to date for companions to very young (5 Myr), very low mass ($<$40
M$_{Jup}$) cluster brown dwarfs.  
We discovered a close companion (0.14$\arcsec$, 20.9$\pm$0.4 AU) 
to the very low mass object SCH
J16091837-20073523.  From spectral deconvolution of 
integrated-light near-IR spectroscopy
of SCH1609-2007 using the SpeX spectrograph (Rayner et al. 2003),
we estimate primary and secondary spectral types of 
M6$\pm$0.5 and M7$\pm$1.0, corresponding to masses of
79$\pm$17 M$_{Jup}$ and 55$\pm$25 M$_{Jup}$ 
at an age of 5 Myr and masses of 84$\pm$15 M$_{Jup}$ and 
60$\pm$25 M$_{Jup}$ at an age of 10 Myr.

For our survey objects with spectral types 
later than M8, we find an upper limit on binary
fraction of $<$9$\%$ (1-$\sigma$) at separations greater than 10 AU. 
As expected from similar mass binaries in the field, 
we find that the binary fraction (10 -- 500 AU separations) 
appears to decrease monotonically with mass 
for young brown dwarfs.
However, while proto-T-dwarfs (M$<$40 M$_{Jup}$) have a similar wide 
(10 -- 500 AU)
binary fraction as field T dwarfs, there exists an anomalous population of 
wide higher mass binaries (0.07 -- 0.1 M$_{\odot}$ primaries,
separations of 10--50 AU)
 at young ages relative to older ages.  

\acknowledgments
The data presented herein were obtained at the
W.M. Keck Observatory, which is operated as a scientific partnership
among the California Institute of Technology, the University of
California and the National Aeronautics and Space Administration. The
Observatory was made possible by the generous financial support of the
W.M. Keck Foundation.  The authors wish to recognize and acknowledge
the very significant cultural role and reverence that the summit of
Mauna Kea has always had within the indigenous Hawaiian community.  We
are most fortunate to have the opportunity to conduct observations
from this mountain.  B.B. was supported by Hubble Fellowship grant
HST-HF-01204.01-A awarded by the Space Telescope Science Institute, 
which is operated by AURA
for NASA, under contract NAS 5-26555.  B.B. would like to acknowledge 
Geoffrey Mathews and Derek Kopon for help with observations and Adam
Kraus and Eric Mamajek for useful discussions.  We thank the 
anonymous referee for useful suggestions which helped improve this work.





\clearpage



\begin{figure}
\epsscale{.80}
\plotone{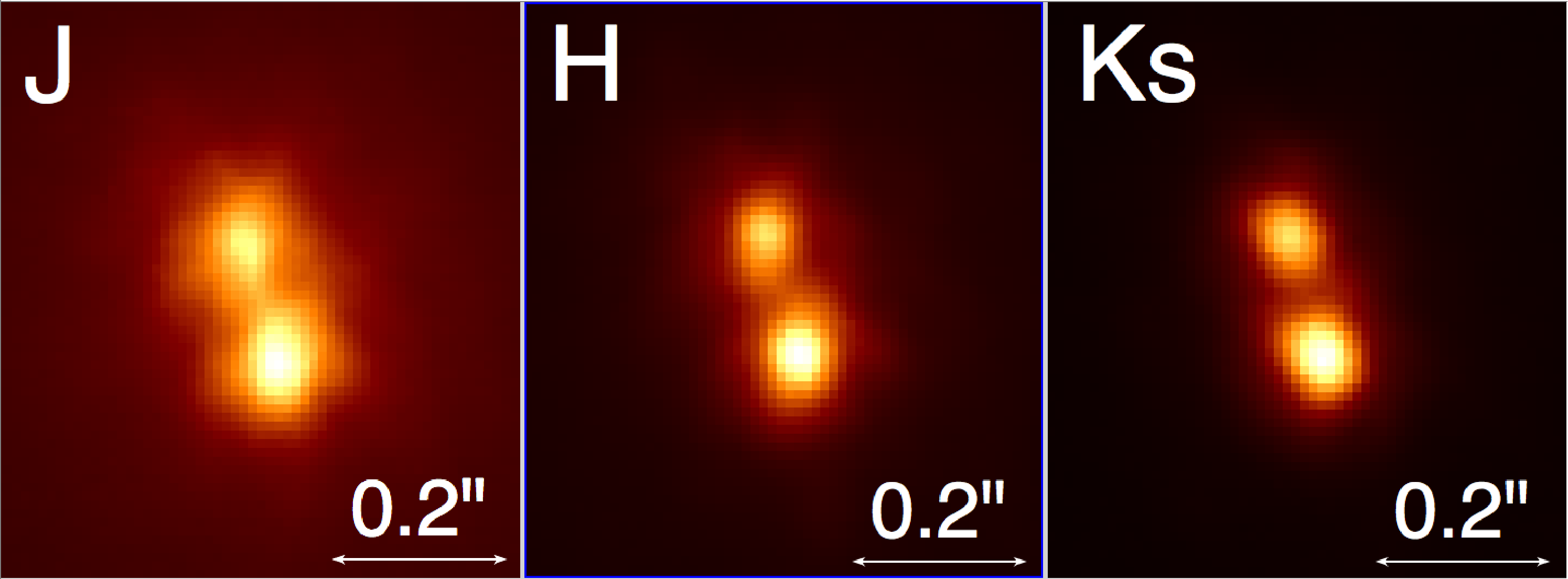}
\caption{Left: {\it J}, {\it H}, 
  and {\it K$_s$}-band images of SCH 16091837-20073523AB
 obtained with NIRC2 and the LGS AO system of the 10m Keck II telescope.
North is up, east is left. Note that primary and companion both appear
slightly elongated in the direction towards the tip-tilt star.
The confirmed companion is at 0.144$\pm$0.002''
  separation and PA=15.87$\pm$0.13$^{\circ}$ with flux ratios of $\Delta${\it J} =
  0.51$\pm$0.09, $\Delta${\it H} = 0.51$\pm$0.03, and $\Delta${\it
    K$_s$} =  0.46$\pm$0.01 mag.  We estimate masses of 47.4$\pm$11.7
  M$_{Jup}$ and 33.5$\pm$6.0 M$_{Jup}$ for primary and companion
  respectively. 
\label{fig:exampleobs}}
\end{figure}

\clearpage

\begin{deluxetable}{lccccccccc}
\tabletypesize{\scriptsize}
\rotate
\tablecaption{Known Young ($<$15 Myr) Very Low Mass
  Mass Binaries (Primary Mass $\leq$0.1 M$_{\odot}$) \label{tab:knownsummary}}
\tablewidth{0pt}
\tablehead{
\colhead{ID} & \colhead{RA} & \colhead{Dec} &
\colhead{SpT1} & \colhead{SpT2} & \colhead{Mass1} & \colhead{Mass2} &
\colhead{Proj. Sep. (\arcsec)} & \colhead{Proj. Sep. (AU)} & \colhead{Ref,Notes}}
\startdata
\multicolumn{10}{c}{Orion (400 pc, $<$1 Myr)} \\ \hline
2MASS J05352184-0546085 & 05:35:21.84 & -05:46:08.5 & M6.5 & M6.5 & 55 M$_{Jup}$
& 35 M$_{Jup}$ & \nodata & 0.04 AU & a,b \\
\hline
\multicolumn{10}{c}{Taurus (140 pc, $<$1 Myr)} \\
\hline
V410-Xray3 & 04:15:01.9\tablenotemark{c} & 28:18:48.\tablenotemark{c} & M6 & M7.7 & 0.093 M$_{\odot}$ &
0.047 M$_{\odot}$ & 0.044$\pm$0.002$\arcsec$ & $\sim$6 AU & d,e \\
MHO-Tau-8 & 04:33:01.1 & 24:21:11.0 & M6 & M6.6 & 0.097 M$_{\odot}$ &
0.073 M$_{\odot}$ & 0.044$\pm$0.008$\arcsec$ & $\sim$6 AU & d,f \\
2MASS J04414489+2301513AB & 04:41:44.89 & 23:01:51.3 & M8.5 & \nodata
& $\sim$20 M$_{Jup}$ & 5-10 M$_{Jup}$ & 1.105$\arcsec$ & 15 AU & g \\
CFHT-Tau 18 & 04:29:21.65 & 27:01:25.95 & M6.0 & \nodata &  0.1 M$_{\odot}$ & 0.06 M$_{\odot}$ &
0.216$\pm$0.002\arcsec & 30.2 AU & h,i \\
CFHT-Tau 7  & 04:32:17.86 & 24:22:14.98 & M6.5 & \nodata & 0.07 M$_{\odot}$ & 0.06 M$_{\odot}$ &
0.224$\pm$0.002\arcsec & 31.4 AU & h,i \\ 
CFHT-Tau 17 & 04:40:01.74 & 25:56:29.23 & M5.75 & \nodata &  0.1 M$_{\odot}$ & 0.06 M$_{\odot}$ &
0.575$\pm$0.002\arcsec & 80.5 AU & h,i \\
FU Tau AB & 04:23:35.39 & 25:03:03.05 & M7.25 & M9.25 & $\sim$0.05
M$_{\odot}$ & $\sim$0.015 M$_{\odot}$ & 5.7$\arcsec$ & 800 AU & j \\
\hline
\multicolumn{10}{c}{Ophiuchus (125 pc, $<$1 Myr) } \\ \hline
Oph 16AB & 16:23:36.09 & -24:02:20.9 & M5$\pm$3 & M5.5$\pm$3 &
$\sim$100 M$_{Jup}$ & $\sim$73 M$_{Jup}$ & 1.7$\arcsec$ & 212$\pm$43
AU & k,l \\
Oph 11AB & 16:22:25.21 & -24:05:13.94 & M9$\pm$0.5 & M9.5$\pm$0.5 &
17$^{+4}_{-5}$ M$_{Jup}$ & 14$^{+6}_{-5}$ & 1.9$\arcsec$ &  243$\pm$55 AU
& k,l,m,n \\ 
\hline
\multicolumn{10}{c}{LkH$\alpha$233 Group (325$^{+72}_{-50}$pc, $\sim$1 Myr) } \\ \hline
2MASS J22344161+4041387 & 22:34:41.61 & 40:41:38.7 & M6 & M6 &
$\sim$0.1 M$_{\odot}$ & $\sim$0.1 M$_{\odot}$ & 0.16$\arcsec$ & 51 AU
& o \\
\hline
\multicolumn{10}{c}{Chamaeleon (160 pc, $<$3 Myr)} \\ \hline
Cha H$\alpha$ 8 & 11:07:47.8 & -77:40:08 & M6.5 & \nodata & 
0.07 - 0.1 M$_{\odot}$ & 30-35 M$_{Jup}$ & \nodata & 1 AU & p,q,r \\
2MASS J11011926-7732383AB & 11:01:19.22 & 77:32:38.60 & M7.25$\pm$0.25
& M8.25$\pm$0.25 & 0.05 M$_{\odot}$ & 0.025 M$_{\odot}$ & 1.44\arcsec
& 240 AU & s \\ 
\hline
\multicolumn{10}{c}{Upper Sco (145 pc, $\sim$5 Myr)} \\ \hline
USco-109AB & 16:01:19.10 & -23:06:38.6 & M6 & M7.5 & 0.07 M$_{\odot}$ & 0.04 M$_{\odot}$ &
0.034$\pm$0.02$\arcsec$ & $\sim$5 AU & t,u \\
USco-66AB & 16:01:49.66 & -23:51:07.4 & M6.0 & M6.0 & 0.07$\pm$0.02
M$_{\odot}$ &  0.07$\pm$0.02 M$_{\odot}$ &
0.07\arcsec & 10.19$\pm$0.07 AU & t,u \\
USco-55AB & 16:02:45.60 & -23:04:49.8 & M5.5 & M6.0 & 0.10$\pm$0.03 M$_{\odot}$ & 0.07$\pm$0.02
M$_{\odot}$ & 0.12\arcsec & 17.63$\pm$0.09 AU & t,u \\
UScoCTIO108AB & 16:05:53.94 & -18:18:42.7 & M7 & M9.5 & 60$\pm$20 M$_{Jup}$ & 14$^{+2}_{-8}$
M$_{Jup}$ & 4.6$\pm$0.1\arcsec & $\sim$670 & v,u \\
\hline
\multicolumn{10}{c}{R Corona Australis ($\sim$130 pc, $\sim$0.5-10 Myr)} \\\hline
DENIS-P J185950.9-370632 & 18:59:50.9 & -37:06:32 & M8$\pm$0.5 &
\nodata & 0.017 M$_{\odot}$ & 0.013 M$_{\odot}$ & 0.06$\arcsec$ & 7.8
AU & w \\
\hline
\multicolumn{10}{c}{TW Hydra ($\sim$30-70 pc, $\sim$12 Myr)} \\\hline
2MASS J1207334−393254 & 12:07:33.40 & −39:32:54.0 & M8 & L5-L9.5 &
$\sim$25 M$_{Jup}$ & 5$\pm$2 M$_{Jup}$ & 0.78\arcsec & 55 AU & x,y \\
\hline
\enddata
\tablenotetext{a}{\citet{sta06}}
\tablenotetext{b}{\citet{sta07}}
\tablenotetext{c}{epoch 1950 coordinates}
\tablenotetext{d}{\citet{kra06}}
\tablenotetext{e}{\citet{str94}}
\tablenotetext{f}{\citet{bri98}}
\tablenotetext{g}{\citet{tod10}}
\tablenotetext{h}{\citet{kon07}}
\tablenotetext{i}{\citet{gui06}}
\tablenotetext{j}{\citet{luh09}}
\tablenotetext{k}{\citet{all06t}}
\tablenotetext{l}{\citet{clo07}}
\tablenotetext{m}{\citet{jay06}}
\tablenotetext{n}{estimated age$\sim$5 Myr}
\tablenotetext{o}{\citet{all09}}
\tablenotetext{p}{\citet{joe06}}
\tablenotetext{q}{\citet{joe07}}
\tablenotetext{r}{\citet{joe10}}
\tablenotetext{s}{\citet{luh04}}
\tablenotetext{t}{\citet{kra05}}
\tablenotetext{u}{\citet{ard00}}
\tablenotetext{v}{\citet{bej08}}
\tablenotetext{w}{\citet{bou04}}
\tablenotetext{x}{\citet{cha05}}
\tablenotetext{y}{\citet{moh07}}
\end{deluxetable}

\begin{deluxetable}{lcccccccc}
\tabletypesize{\scriptsize}
\tablecaption{Objects Observed \label{tab:objects}}
\tablewidth{0pt}
\tablehead{
\colhead{ID} & \colhead{Right Ascension} & \colhead{Declination} & 
\colhead{SpT} & \colhead{J} & \colhead{H} & \colhead{K} & 
\colhead{$\mu_{\alpha} cos \delta$\tablenotemark{a}} & \colhead{$\mu_{\delta}$\tablenotemark{a}}
}
\startdata
\multicolumn{2}{l}{Objects from Lodieu et al. 2008 sample} \\
\hline
USco J155419.99-213543.1 & 15:54:19.99 & -21:35:43.1 & M8 & 14.93 & 14.28 & 13.71 & -14 & -18 \\
USco J160603.75-221930.0 & 16:06:03.75 & -22:19:30.0 & L2 & 15.85 & 15.10 & 14.44 & - & - \\
USco J160606.29-233513.3 & 16:06:06.29 & -23:35:13.3 & L0 & 16.20 & 15.54 & 14.97 & - & - 4 \\
USco J160714.79-232101.2 & 16:07:14.79 & -23:21:01.2 & L0 & 16.56 & 15.83 & 15.07 & - & - 4 \\
USco J160723.82-221102.0 & 16:07:23.82 & -22:11:02.0 & L1 & 15.20 & 14.56 & 14.01 & -11 & -31 \\
USco J160727.82-223904.0 & 16:07:27.82 & -22:39:04.0 & L1 & 16.81 & 16.09 & 15.39 & - & - \\
USco J160737.99-224247.0 & 16:07:37.99 & -22:42:47.0 & L0 & 16.76 & 16.00 & 15.33 & - & - \\
USco J160818.43-223225.0 & 16:08:18.43 & -22:32:25.0 & L0 & 16.01 & 15.44 & 14.70 & - & - \\
USco J160828.47-231510.4 & 16:08:28.47 & -23:15:10.4 & L1 & 15.45 & 14.78 & 14.16 & -12 & -13 \\
USco J160830.49-233511.0 & 16:08:30.49 & -23:35:11.0 & M9 & 14.88 & 14.29 & 13.76 & -5 & -12 \\
USco J160843.44-224516.0 & 16:08:43.44 & -22:45:16.0 & L1 & 18.58 & 17.22 & 16.26 & - & - 12 \\
USco J160847.44-223547.9 & 16:08:47.44 & -22:35:47.9 & M9 & 15.69 & 15.09 & 14.53 & 0 & -20 \\
USco J160918.69-222923.7 & 16:09:18.69 & -22:29:23.7 & L1 & 18.08 & 17.06 & 16.16 & - & - 8 \\
USco J161047.13-223949.4 & 16:10:47.13 & -22:39:49.4 & M9 & 15.26 & 14.57 & 14.01 & -15 & -24 \\
USco J161228.95-215936.1 & 16:12:28.95 & -21:59:36.1 & L1 & 16.41 & 15.56 & 14.79 & - & - \\
USco J161302.32-212428.4 & 16:13:02.32 & -21:24:28.4 & L0 & 17.17 & 16.37 & 15.65 & - & - \\
USco J161441.68-235105.9 & 16:14:41.68 & -23:51:05.9 & L1 & 16.07 & 15.34 & 14.62 & - & - \\
USco J163919.15-253409.9 & 16:39:19.15 & -25:34:09.9 & L1 & 17.20 & 16.39 & 15.61 & -1 & -17 \\ \hline\hline
\multicolumn{1}{l}{Additional Objects} \\
\hline
SCH J16091837-20073523 & 16:09:18.37 & -20:07:35.23 & M7.5 & 13.00 & 12.37 & 12.01 & - & - \\
SCH J16224384-19510575 & 16:22:43.84 & -19:51:05.75 & M8 & 12.35 & 11.61 & 11.15 & - & - \\
\enddata
\tablenotetext{a}{\citet{lod07}}
\end{deluxetable}


\clearpage

\begin{deluxetable}{lccccc}
\tabletypesize{\scriptsize}
\tablecaption{Observations \label{tab:observations}}
\tablewidth{0pt}
\tablehead{
\colhead{ID} & \colhead{Observation Date} & \colhead{Filter} &
\colhead{Exposure Time} & \colhead{Median Strehl} & \colhead{Median FWHM}}
\startdata
\multicolumn{2}{l}{Objects from Lodieu et al. 2008 sample} \\
\hline\hline
USco J155419.99-213543.1 & 2009-05-30 & $K_S$ &  7$\times$60 s & 0.10
& 99 mas \\\hline
USco J160603.75-221930.0 & 2007-07-17 &  $K_S$ &  11$\times$15 s &
0.19 & 66 mas \\
 & & $J$ & 9$\times$30 s & 0.03 & 66 mas \\
 & & $H$ & 9$\times$15 s & 0.06 & 57 mas \\ 
	& 2008-07-27 &  $K_S$ & 11$\times$15 s & 0.31 & 55 mas \\\hline
USco J160606.29-233513.3 & 2009-05-29 &  $K_S$ & 6$\times$60 s & 0.15
& 81 mas \\\hline
USco J160714.79-232101.2 & 2009-05-29 &  $K_S$ & 6$\times$60 s & 0.10
& 80 mas \\
 & & $J$ & 6$\times$60 s & 0.02 & 95 mas \\
 & & $H$ & 6$\times$60 s & 0.05 & 92 mas \\ \hline
USco J160723.82-221102.0 & 2008-07-27 & $K_S$ & 14$\times$15 s & 0.25
& 67 mas \\
 & & $J$ & 12$\times$30 s & 0.03 & 82 mas \\
 & & $H$ & 12$\times$15 s & 0.11 & 66 mas \\ 
  &  2009-05-30 & $K_S$ &  7$\times$60 s & 0.13 & 82 mas \\\hline
USco J160727.82-223904.0 & 2008-07-27 &  $K_S$ & 11$\times$15 s & 0.21
& 62 mas \\
 & & $J$ & 9$\times$30 s & 0.02 & 74 mas\\ \hline
USco J160737.99-224247.0 & 2009-05-29 &  $K_S$ & 6$\times$60 s & 0.11
& 96 mas \\\hline
USco J160818.43-223225.0 & 2009-05-29 &  $K_S$ & 6$\times$60 s & 0.24
& 66 mas \\\hline
USco J160828.47-231510.4 & 2007-07-17 &  $K_S$ & 11$\times$15 s & 0.13
& 82 mas \\\hline
USco J160830.49-233511.0 & 2009-05-30 &  $K_S$ & 7$\times$60 s & 0.06
& 130 mas \\\hline
USco J160843.44-224516.0 & 2009-05-29 &  $K_S$ & 6$\times$60 s & 0.16
& 78 mas \\
 & & $J$ & 6$\times$60 s & 0.02 & 70 mas \\\hline
USco J160847.44-223547.9 & 2007-07-17 &  $K_S$ & 11$\times$15 s & 0.23
& 68 mas \\\hline
USco J160918.69-222923.7 & 2008-07-27 &  $K_S$ & 11$\times$15 s & 0.13
& 67 mas \\\hline
USco J161047.13-223949.4 & 2007-07-17 &  $K_S$ & 11$\times$15 s & 0.19
& 73 mas \\\hline
USco J161228.95-215936.1 & 2008-07-27 &  $K_S$ & 11$\times$15 s & 0.16
& 67 mas \\\hline
USco J161302.32-212428.4 & 2009-05-30 &  $K_S$ & 7$\times$60 s & 0.16
& 79 mas \\\hline
USco J161441.68-235105.9 & 2008-07-27 &  $K_S$ & 11$\times$15 s & 0.16
& 67 mas \\\hline
USco J163919.15-253409.9 & 2008-07-27 &  $K_S$ & 11$\times$15 s & 0.19
& 62 mas \\
 & & $J$ & 9$\times$30 s & 0.02 & 69 mas \\ \hline
\hline
\multicolumn{1}{l}{Additional Objects} \\
\hline\hline
SCH J16091837-20073523 & 2009-06-30 & $K_S$ &  6$\times$20 s & 0.11 &
75 mas \\
& & $J$ & 6$\times$20 s & 0.01 & 80 mas \\
& & $H$ & 6$\times$20 s & 0.04 & 71 mas \\
SCH J16224384-19510575 & 2009-05-30 & $K_S$ &  6$\times$10 s & 0.29 &
59 mas \\\hline
\enddata
\end{deluxetable}

\clearpage



\begin{deluxetable}{lcc}
\tabletypesize{\footnotesize}
\tablecaption{Properties of the SCH1609-2007AB System\label{tab:objphot}}
\tablewidth{0pt}
\tablehead{
\colhead{} & \colhead{Primary} & \colhead{Secondary}}

\startdata

Distance      & \multicolumn{2}{c}{145$\pm$2 pc\tablenotemark{a}} \\
Age           & \multicolumn{2}{c}{5 Myr\tablenotemark{b}} \\
Separation    & \multicolumn{2}{c}{0.144$\pm$0.002\arcsec\ (20.9$\pm$0.4~AU)} \\ 
Position Angle  & \multicolumn{2}{c}{15.87$\pm$0.13\degs} \\
$\Delta{J}$ (mag)                & \nodata  & 0.51$\pm$0.09 \\
$\Delta{H}$ (mag)                & \nodata  & 0.51$\pm$0.03 \\
$\Delta{\Ks}$ (mag)              & \nodata & 0.46$\pm$0.01 \\
$J$ (mag)                        & 13.53$\pm$0.09\tablenotemark{c} & 14.04$\pm$0.09 \\
$H$ (mag)                        & 12.90$\pm$0.04\tablenotemark{c} & 13.41$\pm$0.04 \\
\Ks\ (mag)                       & 12.56$\pm$0.03\tablenotemark{c} &
13.01$\pm$0.03 \\
$J-\Ks$ (mag)                    & 0.97$\pm$0.09 & 1.03$\pm$0.09 \\
$J-H$ (mag)                      & 0.63$\pm$0.10 & 0.63$\pm$0.10 \\
$H-\Ks$ (mag)                    & 0.34$\pm$0.05 & 0.40$\pm$0.05 \\
Log$\frac{L}{L_{\odot}}$         &  -2.04$\pm$0.12 & -2.23$\pm$0.12 \\
Spectral Type          & M7$\pm$0.5  & M6$\pm$1.0 \\
$T_{eff}$    & 2990$\pm$60 K & 2850$\pm$170 K \\
Estimated Mass (5 Myr) & 79$\pm$17 M$_{Jup}$ & 55$\pm$25 M$_{Jup}$ \\ 
Estimated Mass (10 Myr) &  84$\pm$15 M$_{Jup}$ & 60$\pm$25 M$_{Jup}$ \\
\enddata
\tablenotetext{a}{\citet{pre02}}
\tablenotetext{b}{\citet{pre99}}
\tablenotetext{c}{from 2MASS}
\end{deluxetable}

\begin{deluxetable}{lccccccc}
\tabletypesize{\scriptsize}
\tablecaption{Measured $K_s$ Contrast and Minimum Detectable Mass Ratios\label{tab:cont_q}}
\tablewidth{0pt}
\tablehead{
\colhead{ID} & \colhead{Observation Date} & 
\colhead{$\Delta$mag(0.07\arcsec)} & \colhead{q(0.07\arcsec)} &
\colhead{$\Delta$mag(0.2\arcsec)} & \colhead{q(0.2\arcsec)} &
\colhead{$\Delta$mag(0.5\arcsec)} & \colhead{q(0.5\arcsec)}}
\startdata
USco J155419.99-213543.1   &     2009-05-30   &   0.41  &    0.87   &      3.67  &    0.24  &    5.54  &    0.11 \\
USco J160603.75-221930.0    &     2008-07-27   &   1.83  &    0.46   &5.01   &   0.13   &   5.88  &    0.08 \\  
USco J160603.75-221930.0    &     2007-07-17   &   1.17  &    0.59   &   4.40   &   0.16   &   5.42  &    0.10 \\
USco J160606.29-233513.3   &  2009-05-29  &    0.65  &    0.71  &    3.79  &    0.20  &      4.84  &      0.13 \\
USco J160714.79-232101.2       &  2009-05-29  &    0.70  &    0.71  &    3.89  &    0.19  &      4.66  &      0.14 \\
USco J160723.82-221102.0        &  2008-07-27  &    1.23  &    0.64  &    4.95  &    0.14  &      5.98  &      0.08 \\
USco J160723.82-221102.0         &  2009-05-30  &    0.62  &    0.80  &    3.74  &    0.23  &      5.53  &      0.11 \\
USco J160727.82-223904.0         &  2008-07-27  &    1.36  &   0.54   &   4.28   &   0.16   &     4.55  &      0.14 \\
USco J160737.99-224247.0       &  2009-05-29  &    0.42  &    0.86  &    3.45  &    0.23   &     4.22  &      0.16 \\
USco J160818.43-223225.0      &  2009-05-29  &    1.16  &    0.58  &    4.53  &    0.15    &    5.61  &      0.08 \\
USco J160828.47-231510.4       &  2007-07-17  &    0.59  &    0.80  &    4.14  &    0.19    &    5.26  &      0.12 \\
USco J160830.49-233511.0       &  2009-05-30  &    0.14  &    0.95  &    2.91  &    0.32  &      5.11  &      0.13 \\
USco J160843.44-224516.0        &  2009-05-29  &    0.78  &    0.72  &    3.20  &    0.27  &      3.47  &      0.23 \\
USco J160847.44-223547.9         &  2007-07-17  &    1.23  &    0.56  &    4.68  &    0.14  &      5.51  &      0.10 \\
USco J160918.69-222923.7       &  2008-07-27  &    1.11  &    0.63  &    3.46  &    0.23  &      3.66  &      0.21 \\
USco J161047.13-223949.4        &  2007-07-17  &    0.82  &    0.75  &    4.49  &    0.17  &      5.66  &      0.10 \\
USco J161228.95-215936.1        &  2008-07-27  &    1.08  &    0.60  &    4.23  &    0.17  &      4.78  &      0.13 \\
USco J161302.32-212428.4         &  2009-05-30  &    0.69  &    0.76  &    3.91  &    0.18   &     4.70  &      0.11 \\
USco J161441.68-235105.9        &  2008-07-27  &    1.48  &    0.52  &      4.52   &   0.15   &   5.18  &    0.11 \\
USco J163919.15-253409.9    &  2008-07-27  &    1.38  &    0.56  &      3.98  &    0.18  &    4.36  &    0.14 \\
\enddata
\end{deluxetable}

\begin{deluxetable}{ccccc}
\tabletypesize{\scriptsize}
\tablecaption{Statistical Sample Comparison as a Function of Mass\label{tab:tests1}}
\tablewidth{0pt}
\tablehead{
\colhead{Sample 1} & \colhead{Sample 2} & \colhead{Likelihood} &
\colhead{1-$\sigma$ CI on $\delta=\epsilon_{bin1} - \epsilon_{bin2}$} &
\colhead{2-$\sigma$ CI on $\delta=\epsilon_{bin1} - \epsilon_{bin2}$}}
\startdata
\multicolumn{5}{c}{Upper Sco Mass Comparison (10 -- 1000 AU separations) } \\ \hline
$<$0.04 M$_{\odot}$ & 0.04--0.1 M$_{\sun}$ & & & \\
0 / 18,  $<$9$\%$ & 2 / 12, 17$_{-6}^{+15}$\% & 0.15 & -0.28, -0.15 & -0.42, 0.04\\\hline
\multicolumn{5}{c}{Upper Sco, Taurus, and Chamaeleon Mass Comparison
  (10 -- 1000 AU separations) } \\ \hline
$<$0.04 M$_{\odot}$ & 0.07--0.1 M$_{\sun}$ & & &  \\
0 / 25, $<$7$\%$ & 6 / 23, 26$_{-7}^{+11}$\% & 0.01 & -0.34, -0.15 &  -0.44, -0.07 \\ \hline 
$<$0.04 M$_{\odot}$ & 0.04--0.07 M$_{\sun}$ & & & \\
0 / 25, $<$7$\%$ & 0 / 18, $<$9$\%$ & 1.0 & -0.06, 0.04 & -0.15, 0.10 \\ \hline
0.04--0.07 M$_{\sun}$ & 0.07--0.1 M$_{\sun}$ & & & \\
0 / 18, $<$9$\%$ &  6 / 23, 26$_{-7}^{+11}$\% & 0.03 & -0.33, -0.13&-0.43, -0.04 \\\hline
\multicolumn{5}{c}{Upper Sco, Taurus, Chamaeleon, and IC 348 Mass
  Comparison (30 -- 1000 AU separations)} \\ \hline
$<$0.04 M$_{\odot}$ & 0.07--0.1 M$_{\sun}$ & & &  \\
0 / 36, $<$4.8$\%$ & 3 / 43, 7$_{-2}^{+6}$\% & 0.25 & -0.11, -0.02 &-0.17, 0.03 \\ \hline 
$<$0.04 M$_{\odot}$ & 0.04--0.07 M$_{\sun}$ & &  & \\
0 / 36, $<$4.8$\%$ & 0 / 20, $<$8$\%$ & 1.0 & -0.06, 0.02 & -0.14, 0.07 \\ \hline
0.04--0.07 M$_{\sun}$ & 0.07--0.1 M$_{\sun}$ & & & \\
0 / 20, $<$8$\%$ &  3 / 43, 7$_{-2}^{+6}$\% & 0.54 & -0.10, 0.01 & -0.16, 0.09 \\\hline
\enddata
\end{deluxetable}

\begin{deluxetable}{ccccc}
\tabletypesize{\scriptsize}
\tablecaption{Statistical Sample Comparison as a Function of Age\label{tab:tests2}}
\tablewidth{0pt}
\tablehead{
\colhead{Sample 1} & \colhead{Sample 2} & \colhead{Likelihood} &
\colhead{1-$\sigma$ CI on $\delta=\epsilon_{bin1} - \epsilon_{bin2}$} &
\colhead{2-$\sigma$ CI on $\delta=\epsilon_{bin1} - \epsilon_{bin2}$}}
\startdata
\multicolumn{5}{c}{Upper Sco vs. Field (10 -- 1000 AU separations) } \\ \hline
$<$0.04 M$_{\odot}$ BDs in Upper Sco & Field T Dwarfs & & & \\
0 / 18, $<$9$\%$ & 0 / 32, $<$5$\%$ & 1.0 & -0.02, 0.07 & -0.08, 0.15\\\hline
\multicolumn{5}{c}{Upper Sco, Taurus, and Chamaeleon vs. Field (10 --
  1000 AU separations)} \\ \hline
0.07--0.1 M$_{\odot}$ Cluster BDs & Field M and L Dwarfs & & & \\
6 / 23, 26$_{-7}^{+11}$\% & 1 / 39, 2.6$_{-0.1}^{+5.4}$\% & 0.01 & 0.14, 0.32 & 0.06, 0.43 \\\hline
\multicolumn{5}{c}{Upper Sco, Taurus, Chamaeleon, and IC 348 vs. Field (30 -- 1000 AU separations) } \\ \hline
0.07--0.1 M$_{\odot}$ Cluster BDs & Field M and L Dwarfs & & & \\
3 / 43, 7$_{-2}^{+6}$\% & 0 / 39, $<$4.4$\%$ & 0.24 & 0.02, 0.11 & -0.02, 0.17 \\\hline
\enddata
\end{deluxetable}
\end{document}